\documentclass[12pt%
,final%
]{nuthesis}	%

\usepackage[pass]{geometry}

\usepackage{xcolor}
\usepackage{siunitx}
\usepackage{soul}
\usepackage{amssymb}
\usepackage{booktabs} 
\usepackage{enumitem}

\usepackage[nocompress]{cite}

\usepackage{algpseudocodex}
\usepackage[chapter]{algorithm}

\newcounter{algoline}[algorithm]

\makeatletter
\patchcmd{\ALG@step}{\addtocounter{ALG@line}{1}}{\stepcounter{ALG@line}\refstepcounter{algoline}}{}{}

\makeatother

\makeatletter

\renewcommand*{\listof}[2]{%
  \@ifundefined{ext@#1}{\float@error{#1}}{%
    \@namedef{l@#1}{\l@table}%
    \begingroup\setlength{\parskip}{\z@}%
      \@starttoc{\@nameuse{ext@#1}}{#2}%
    \endgroup}}

\makeatother

\usepackage{graphicx}
\graphicspath{ {./figures/} }

\usepackage{subcaption}

\usepackage{listings}
\usepackage{fancyvrb}

\makeatletter
\let\l@lstlisting=\l@table
\makeatother

\definecolor{codegreen}{rgb}{0,0.5,0}
\definecolor{codegray}{rgb}{0.5,0.5,0.5}
\definecolor{codepurple}{rgb}{0.58,0,0.82}
\definecolor{codeblue}{rgb}{0.13,0.13,1}
\definecolor{codered}{rgb}{0.9,0,0}

\definecolor{backcolour}{rgb}{1.0,1.0,1.0}

\lstdefinestyle{mystyle}{
    backgroundcolor=\color{backcolour},   
    commentstyle=\color{codegreen},
    keywordstyle=\color{codeblue},
    keywordstyle=[2]{\color{codepurple}},
    keywordstyle=[3]{\color{magenta}},
    numberstyle=\tiny\color{codegray},
    stringstyle=\color{codered},
    basicstyle=\ttfamily\footnotesize,
    breakatwhitespace=false,         
    breaklines=true,                 
    captionpos=b,                    
    keepspaces=true,                 
    numbers=left,                    
    numbersep=2pt,                  
    showspaces=false,                
    showstringspaces=false,
    showtabs=false,                  
    tabsize=4,
    frame=lines,
    framexleftmargin=5mm
}

\usepackage[left,running]{lineno}

\makeatletter
\renewcommand\@pnumwidth{1em} %
\patchcmd{\@tocline}
  {\hfil}
  {\leaders\hbox{\,.\,}\hfil}
  {}{}
\makeatother

\author{Nikola Vuk Maruszewski}

\title{Improved Prefetching Techniques for Linked Data Structures}

\degree{MASTER OF SCIENCE}  %

\field{Computer Engineering}            %

\graduationmonth{June}         %

\graduationyear{2025}          %


\begin{document}

\abstractText{%
With ever-increasing main memory stall times, we need novel techniques to reduce effective memory access latencies. Prefetching has been shown to be an effective solution, especially with contiguous data structures that follow the traditional principles of spatial and temporal locality. However, on linked data structures---made up of many nodes linked together with pointers---typical prefetchers struggle, failing to predict accesses as elements are arbitrarily scattered throughout memory and access patters are arbitrarily complex and hence difficult to predict. To remedy these issues, we introduce \textsc{Linkey}, a novel prefetcher that utilizes hints from the programmer/compiler to cache layout information and accurately prefetch linked data structures. \textsc{Linkey} obtains substantial performance improvements over a striding baseline. We achieve a geomean 13\% reduction in miss rate with a maximum improvement of 58.8\%, and a 65.4\% geomean increase in accuracy, with many benchmarks improving from 0\%. On benchmarks where \textsc{Linkey} is applicable, we observe a geomean IPC improvement of 1.40\%, up to 12.1\%.%
}

\frontmatter		%

\maketitle		%

\copyrightpage		%

\abstract %

\acknowledgements

First, I would of course like to thank my advisor, Prof. Nikos Hardavellas. I have had the privilege to work with him for the past three years; his guidance and support has helped me become the researcher I am today. I'd also like to thank the other members of my thesis committee, Prof. Peter Dinda and Prof. Russ Joseph, for taking the time to review this dissertation and attend my defense. 

I'd like to thank Karl Hallsby for his help in drafting and reviewing this thesis. Tommy M\textsuperscript{c}Michen's advice on data structures was also invaluable for designing benchmarks. Additionally, I'd like to thank Nick Wanninger and Steve Ewald for helping make the benchmarks more representative, and Atmn Patel for his advice on hardware resources. Finally, I must thank Joe Maruszewski for the information on octrees and their traversal patterns in Computational Fluid Dynamics programs.

I would also like to acknowledge my prior collaborators, who have helped me grow as a researcher and hone my skills. A big thank you to Prof. Kate Smith, Jessica Jeng, Michael Gavrincea, and Connor Selna. 

Last but not least, I must thank my family for all they have done for me, and their continuing support throughout my academic journey. We are all products of our environment, and I thank my family for encouraging me to push myself and helping me become the person I am today.

\listofabbreviations 

\begin{description}
    \item[AT] Address Table
    \item[BFQ] Backup Fetch Queue
    \item[BFS] Breadth-First Search
    \item[BST] Binary Search Tree
    \item[CAT] Child Association Table
    \item[CDP] Content Directed Prefetcher~\cite{cooksey_StatelessContentdirectedData_2002}
    \item[DFS] Depth-First Search
    \item[LDS] Linked Data Structure
\end{description}

\dedication{%
    To my grandparents, Milan and Ljubica Egelja. Without your teachings and support, this thesis would have never been written.
}

\clearpage\phantomsection %
\setcounter{tocdepth}{2}
\tableofcontents	%

\clearpage\phantomsection %
\listoftables		%

\clearpage\phantomsection %
\listoffigures		%

\clearpage\phantomsection %
\listofalgorithms

\clearpage\phantomsection %
\makeatletter
\begingroup
\@starttoc{lol}{List of Listings}
\endgroup
\makeatother

\mainmatter             %

\chapter{Introduction}
\label{chap:intro}

Effective main memory access times have become the main bottleneck in modern high-performance computing systems. While processor performance has followed Moore's law, doubling every 1.5 years, DRAM speeds have failed to keep up, with access latencies growing to at least \textbf{300 cycles} on modern processors~\cite{mori_LocalizingTagComparisons_2024}. As this processor-memory performance gap widens, new developments in processor technology will go unnoticed if memory stall times cannot be reduced. We must develop new techniques to hide this latency if we wish to maximize the performance of our systems.

Prefetching has been shown to be an effective solution for reducing the effective memory access latencies in modern high-performance processors~\cite{smith_CacheMemories_1982}. By predicting the next memory access and ensuring the data is in the \mbox{L1-D\$} before the processor issues the corresponding memory operation, prefetchers can, in theory, completely hide the latencies of the L2\$ and below. With contiguous data structures, such as arrays or hashtables, this technique has been incredibly effective, even with complex access patterns~\cite{mittal_SurveyRecentPrefetching_2017}.

Unfortunately, modern prefetching techniques struggle when presented with linked data structures (LDSs), such as linked lists, trees, and graphs. As the nodes of these data structures are arbitrarily distributed throughout memory and traversed through pointer chasing, there is no inter-element spatial locality to take advantage of like in contiguous data structures. Additionally, as access patterns to these objects are often unpredictable (e.g., searching for an arbitrary key in a binary search tree), prefetching techniques that take advantage of temporal locality such as correlation~\cite{mittal_SurveyRecentPrefetching_2017, somogyi_SpatiotemporalMemoryStreaming_2009, wenisch_TemporalStreamingShared_2005, joseph_PrefetchingUsingMarkov_1997} often fail as well.

One might wonder, if these data structures are so detrimental to performance, are they really so commonly used to warrant special optimization? Programmers are taught to maximize spatial and temporal locality, if they wish to maximize performance, why would they use a data structure with poor locality? A map can be implemented as either a hashtable or a binary tree, it seems like the hashtable should be the clear choice for speed.

There are conditions where linked data structures are faster than their array-based counterparts, or even required to satisfy the requirements of a program. As they are constructed of nodes connected by pointers, it is rather trivial to make modifications to LDSs, irrespective of the location of the modified element. For example, an insertion into a linked list takes $\mathcal{O}(1)$ time, regardless of whether the location is the start, end, or middle, while a dynamic array only has $\mathcal{O}(1)$ insertion at the end. Linked data structures are also more memory efficient, as only the exact number of nodes needed to hold the data need be allocated, compared to array-based structures where an \textit{exponentially growing} over-allocation is often required for good amortized time complexity. Some LDSs even provide guarantees not found elsewhere, such as binary search trees~(BSTs) that offer both fast lookup and ordering.

It is no surprise then that linked data structures are incredibly common across many applications. The Linux kernel~\cite{torvalds_KernelGitTorvalds_2025} makes heavy use of (intrusive) linked lists for tracking free objects, along with balanced BSTs for scheduling and I/O\@. Other types of trees are also used for scientific computing~\cite{khokhlov_FullyThreadedTree_1998}, database indexes~\cite{comer_UbiquitousBTree_1979}, fast text search~\cite{aho_EfficientStringMatching_1975}, and bioinformatics~\cite{martinez-prieto_PracticalCompressedString_2016}. Finally, sparse graphs, implemented using adjacency lists, are widely used for social networks analysis, web crawling, and biology~\cite{besta_SurveyTaxonomyLossless_2019}.

Clearly, the importance and prevalence of these linked data structures justifies targeted optimization for their specific access patterns. Unfortunately, as stated previously, most prefetching techniques optimize for traditional locality, which often fails when presented with pointer-chasing applications. To be most effective, we must take advantage of \textit{reference locality}---the idea that the set of objects referenced by a given object is typically unchanging~\cite{ayers_ClassifyingMemoryAccess_2020}.

Content Directed Prefetchers~(CDPs), as introduced in~\cite{cooksey_StatelessContentdirectedData_2002}, attempt to optimize for reference locality. Their distinguishing feature is the use of memory \textit{responses}, in addition to request addresses, to search for possible ``recursive'' pointers (i.e., pointers to nodes of the same data structure) and issue prefetches. However, as many CDPs strive for a hardware-only implementation~\cite{ebrahimi_TechniquesBandwidthefficientPrefetching_2009, wang_GuidedRegionPrefetching_2003, liu_SemanticsAwareTimelyPrefetching_2010}, they must speculate if a value is a pointer, leading to cache pollution and even security vulnerabilities~\cite{vicarte_AuguryUsingData_2022, chen_GoFetchBreakingConstantTime_2024}. Other implementations rely heavily on software/profiling~\cite{burcea_PointyHybridPointer_2012, zhang_PDGPrefetcherDynamic_2024, wang_GuidedRegionPrefetching_2003, zhang_RPG2RobustProfileGuided_2024} or require substantial hardware resources~\cite{hughes_PrefetchingLinkedData_2000, yang_ProgrammableMemoryHierarchy_2002, yang_PushVsPull_2000, huang_PerformanceAnalysisPrefetching_2009, huang_EstimatingEffectivePrefetch_2012, huang_PerformanceOptimizationThreaded_2012}.

To remedy these issues and improve performance in pointer-chasing applications, we introduce a novel prefetching technique for linked data structures, which we have termed \textsc{Linkey}. We take a hybrid approach with hardware-software co-design to provide high-performance while eliminating speculation on whether a value is a pointer. Metadata specifying basic LDS node layout information is passed down to the hardware, alleviating the need to perform shape analysis in silicon. \textsc{Linkey} is also provided with at least one known memory location in the data structure (e.g., the root), from which it learns the shape of the entire structure. With these simple bits of information, \textsc{Linkey} can continue on to effectively prefetch complex linked data structures.

\section{Contributions}

\noindent Our contributions are as follows:
\begin{enumerate}
    \item We introduce \textsc{Linkey}, a novel prefetching technique for linked data structures. With only a tiny amount of information passed down from the software, \textsc{Linkey} can effectively issue prefetch requests for pointer-chasing access patterns. \textsc{Linkey} is also very adaptable, supporting data structures with varying numbers of linking (i.e., recursive) pointers.
    \item We evaluate \textsc{Linkey} against traditional prefetcher technologies on a representative set of LDS benchmarks, showing substantial performance improvements over a striding baseline. We achieve a geomean 13\% reduction in miss rate with a maximum improvement of 58.8\%, and a 65.4\% geomean increase in accuracy, with many benchmarks improving from 0\%. On benchmarks where \textsc{Linkey} is applicable, we observe a geomean IPC improvement of 1.40\%, up to 12.1\%.
\end{enumerate}

\section{Thesis Organization}

The remainder of this thesis is structured as follows: \autoref{chap:background} gives background information on modern prefetching technologies and further motivates the design of \textsc{Linkey}. \autoref{chap:design} delves into the design and implementation of \textsc{Linkey}. Next, \autoref{chap:expirements} benchmarks \textsc{Linkey} against an array of other prefetching techniques. In \autoref{chap:related_works}, we discuss other related works for prefetching pointer-chasing access patterns. Finally, we conclude and discuss future work in \autoref{chap:conclusion}.

\chapter{Background and Motivation}
\label{chap:background}

Linked data structures are a critical component of many algorithms and software programs. However, as they do not consistently follow the traditional principles of temporal and spatial locality, predicting accesses is a major challenge. Prefetchers have attempted to improve accuracy using software, hardware, and hybrid approaches with varying degrees of success and practicality. However, with the current trends in memory system design, even the best of these approaches will still struggle when presented with linked data structures.

\section{Linked Data Structures}

Linked data structures~(LDSs) are one of the two main categories of data structures, the other being contiguous data structures. The main difference between them is how elements are found and traversed. A contiguous data structure is allocated as a single large block of memory, which enables fast random access times but means modifications can be slow if a re-allocation is required. Contiguous data structures are also often over-allocated to provide fast amortized time complexities, leading to memory overheads, as the size of the over-allocation must grow \textit{exponentially} for the amortization to work.

On the other hand, LDSs are built from many small nodes \textbf{linked} together with pointers. This means each allocation happens individually, and nodes can be arbitrarily scattered throughout memory. This hurts random access times, as one may have to traverse a long pointer chain to access a node. Modifications, on the other hand, are significantly simpler, as re-allocations and movements of large blocks of memory are not required---only a few pointers must be updated. Additionally, LDSs are space efficient---no over-allocations are \textit{required} for their complexity guarantees, and any over-allocation for low-level performance optimizations (such as a node pool) only needs to grow \textit{linearly}.

\begin{figure}[h]
    \centering
    \begin{subfigure}[h]{0.24\textwidth}
        \centering
        \includegraphics[width=\textwidth]{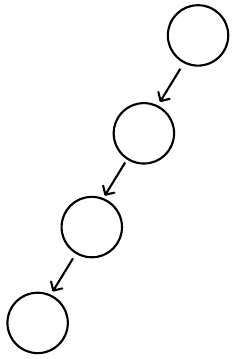}
        \caption{Linked List}
        \label{fig:types_of_lds:ll}
    \end{subfigure}
    ~
    \begin{subfigure}[h]{0.24\textwidth}
        \centering
        \includegraphics[width=\textwidth]{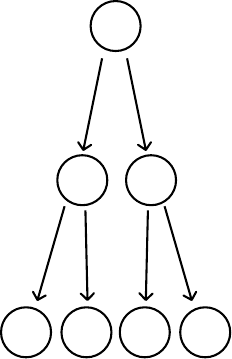}
        \caption{Binary Tree}
        \label{fig:types_of_lds:bintree}
    \end{subfigure}
    ~
    \begin{subfigure}[h]{0.24\textwidth}
        \centering
        \includegraphics[width=\textwidth]{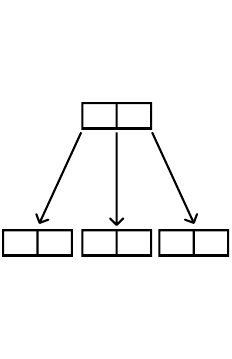}
        \caption{B-Tree}
        \label{fig:types_of_lds:b-tree}
    \end{subfigure}
    ~
    \begin{subfigure}[h]{0.24\textwidth}
        \centering
        \includegraphics[width=\textwidth]{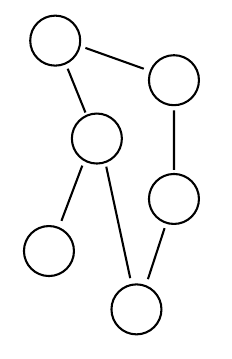}
        \caption{Adjacency List}
        \label{fig:types_of_lds:graph}
    \end{subfigure}
    \caption[Types of linked data structures]{Common types of linked data structures. Circles and rectangles represent nodes, the arrows represent pointers (i.e., the links) between nodes in the linked data structure.}
    \label{fig:types_of_lds}
\end{figure}

Due to these benefits, there are many applications that make use of linked data structures. Memory allocators in glibc and the Linux kernel utilize linked lists (shown in \autoref{fig:types_of_lds:ll}) to track free segments of memory. Self-balancing binary search trees (\autoref{fig:types_of_lds:bintree}), such as red-black trees~\cite{guibas_DichromaticFrameworkBalanced_1978}, are used in Linux's Completely Fair Scheduler for ordering processes, and in \texttt{epoll(7)} for tracking file descriptors~\cite{torvalds_KernelGitTorvalds_2025}. Octrees (8-ary trees) have uses in 3D graphics~\cite{meagher_GeometricModelingUsing_1982} and scientific computing applications such as computational fluid dynamics~\cite{khokhlov_FullyThreadedTree_1998} that partition 3D space. Additionally, tries (i.e., prefix trees)~\cite{delabriandais_FileSearchingUsing_1959} have been used for fast text search~\cite{aho_EfficientStringMatching_1975}, web server routing~\cite{nouvertne_Litestar_2024}, and bioinformatics~\cite{martinez-prieto_PracticalCompressedString_2016}. In database systems, B-Trees (\autoref{fig:types_of_lds:b-tree}) and their variations have been heavily used, mainly for the implementation of table indexes~\cite{comer_UbiquitousBTree_1979}. Finally, sparse graphs are most often implemented using adjacency lists (\autoref{fig:types_of_lds:graph}), with applications in social networks analysis, web crawling, and biology~\cite{besta_SurveyTaxonomyLossless_2019}.

In many of these applications, an LDS is used simply because there is no other alternative. The Linux kernel scheduler must have predictable runtime, thus a hashmap or array-based binary heap that may need to perform a lengthy resize would be unsuitable. Database indexes require a data structure that supports both point and range lookups, a feature unique to the B-Tree family; the use of two separate indexes for each feature would be untenable due to storage overheads. Most of all, the incredibly large sparse graphs used in social network analysis must be implemented using an adjacency list---any other implementation would consume orders of magnitude more memory. Clearly, LDSs are important enough to warrant targeted optimizations.

\section{Principle of Locality}

Code executed on modern CPUs tends to follow the principles of spatial and temporal locality: nearby objects are often accessed together, and frequently accessed objects will likely be accessed again in the future. Modern systems stacks optimize on this principle, and thus programmers optimize code by improving locality, creating a virtuous cycle. Prefetchers notably employ locality to perform shape analysis; they speculate on the layout of an object in memory using it and issue requests accordingly. This technique is often extremely effective, as contiguous data structures are very common (vectors, hashmaps, etc.) and some pointer-based access patters are frequently repeated (e.g., virtual function calls).

Unfortunately, linked data structures break this paradigm. As each node is typically allocated individually, the user has no control over its placement in memory; this arbitrary distribution of nodes means any spatial locality-based prefetcher struggles when presented with an LDS, as the location of one node gives no indication of the location of others. While there is some temporal locality in access patterns, and certain nodes can  be hotter than others depending on the type of traversal, it is still just an approximation. A prefetcher for linked data structures should instead consider \textbf{reference locality}: the principle that the set of nodes \textit{referenced} by a given node tends not to change. This implies child pointers of a node are relatively constant, which also explains why temporal locality can sometimes serve as a proxy for reference locality when access patterns are crystalline (i.e., not noisy) and repetitive.

\section{Prefetcher Implementations}

Prefetchers appeared in research as early as 1982~\cite{smith_CacheMemories_1982}, and have grown more and more complex since. Implementations have been done in either software or hardware, primarily focusing on the memory operation address stream. Only recently have new designs bridged the gap between software and hardware, or started using memory \textit{responses} to inform future prefetches.

\subsection{Software Implementations}

A common method of implementing prefetching is by using a compiler to insert speculative load (i.e., prefetch) instructions into a program. When a processor hits these instructions, it will issue a load, but suppress any errors due to memory protection violations or invalid loads. The result of the load is also dropped, as the only purpose is to bring data into the cache. From a hardware perspective, this is trivial to implement---only a slightly modified load instruction is needed---and it allows prefetching schemes to be tailored to the program executed. An example of software prefetching for linked lists and vectors is shown in \autoref{lst:sw_prefetching}.

\begin{lstlisting}[
    language=C,
    float,floatplacement={h!tbp},
    label={lst:sw_prefetching},
    caption={[Software prefetching example]{Software prefetching example for vectors and linked lists.}}
]
void vec_work(int arr*, size_t N) {
    for (int i = 0; i < N; ++i) {
        <@{\color{magenta}prefetch}@>(arr[i + 1]);
        work(arr[i]);
    }
}

void ll_work(node_t* n) {
    for (node_t* cur = n; cur != NULL;) {
        <@{\color{magenta}prefetch}@>(cur->next);
        work(cur->data);
        cur = cur->next;
    }
}
\end{lstlisting}

Unfortunately, the separation of the compiler and runtime leads to issues with pure software implementations. As the compiler is not fully aware of microarchitecural details and system configurations, such as memory speeds, it must use an (inherently flawed) cost model to determine prefetch depth and scheduling. Additionally, when prefetching linked data structures, the compiler must either rely on alias/shape analysis or profiling to decide what to prefetch~\cite{ayers_ClassifyingMemoryAccess_2020, zhang_RPG2RobustProfileGuided_2024, wang_GuidedRegionPrefetching_2003}; the former is intractable as it maps to the halting problem~\cite{ramalingam_UndecidabilityAliasing_1994}, while to be accurate the latter requires coverage of all program inputs and/or extremely low overheads to run online, which is very difficult to accomplish in practice. Finally, the insertion of additional instructions into hot loops will raise code size, increasing \mbox{L1-I\$} pressure and hurting performance.

\subsection{Hardware Implementations}

Hardware has been the primary implementation of prefetching algorithms for many years. The simplest ``Next-K'' prefetchers bring in the next $K$ lines after the current miss, while striding prefetchers fetch the $n$-th next cache line after the current one. Both strategies optimize for contiguous, sequential traversals such as vectors~\cite{mittal_SurveyRecentPrefetching_2017, smith_CacheMemories_1982}. More advanced correlation-based prefetchers~\cite{mittal_SurveyRecentPrefetching_2017, joseph_PrefetchingUsingMarkov_1997, somogyi_SpatiotemporalMemoryStreaming_2009, wenisch_TemporalStreamingShared_2005} observe repeated patterns of accesses at a constant offset (e.g., $A$, $A+8$, $B$, $B+8$, $C$, $C+8$); such designs have seen greater success with linked data structures and other pointer-chasing applications. Even more complex designs have turned to implementing entire neural networks in hardware, which are in theory adaptable across many applications~\cite{peled_NeuralNetworkPrefetcher_2019}.

Hardware prefetcher designs have been previously created with pointer-chasing access patterns in mind. Some techniques compute and store jump pointers during accesses to an LDS, allowing ``long-distance'' prefetches without traversing a pointer chain~\cite{luk_AutomaticCompilerinsertedPrefetching_1999, roth_EffectiveJumppointerPrefetching_1999}. Other designs extract a \textit{prefetch kernel} from the application and schedule it to run on a custom prefetch engine~\cite{yang_ProgrammableMemoryHierarchy_2002, hughes_PrefetchingLinkedData_2000, yang_PushVsPull_2000},  a second SMT thread~\cite{huang_EstimatingEffectivePrefetch_2012, huang_PerformanceAnalysisPrefetching_2009, huang_PerformanceOptimizationThreaded_2012}, or insert it directly into the application~\cite{sankaranarayanan_HelperThreadsCustomized_2020}. Techniques for improving the effective latency of indirect memory accesses use pointer caches for load address prediction~\cite{collins_PointerCacheAssisted_2002} or track memory access patterns to predict the indirection function and issue prefetches accordingly~\cite{yu_IMPIndirectMemory_2015, xue_TycheEfficientGeneral_2024}. Finally, more advanced hardware-only designs attempt to infer an LDS's semantic structure (i.e., shape) from the memory request address stream~\cite{liu_SemanticsAwareTimelyPrefetching_2010}.

\subsubsection{Content Directed Prefetchers}

This new class of prefetchers, introduced in~\cite{cooksey_StatelessContentdirectedData_2002}, not only examines the address stream from the processor, but also the \textbf{responses} from the memory system. Cooskey et al.\ specifically attempt to find pointers in cache lines returned from memory and issue prefetches for those addresses, in the hopes that they are child pointers in an LDS\@. This technique was further enhanced in~\cite{ebrahimi_TechniquesBandwidthefficientPrefetching_2009} to attempt to reduce prefetch misses. Apple also recently used this technique in their M-series of processors; unfortunately, it was shown in~\cite{chen_GoFetchBreakingConstantTime_2024} and~\cite{vicarte_AuguryUsingData_2022} that speculating if a value is a pointer opens up many security vulnerabilities. This is in addition to cache pollution caused by the inability to identify if a pointer is referencing a child node or simply points to other data that won't be needed in the traversal (such as parent nodes, metadata, etc.).

\subsection{Hybrid Implementations}

Recent prefetcher designs have attempted to cross the boundary between hardware and software for further optimization opportunities, combining high-level information easily determined by the compiler, such as object sizes and types, with the low-level runtime information readily available to the CPU, such as pointer values. Some of these techniques try to inform the hardware about every object pointer in the application~\cite{burcea_PointyHybridPointer_2012} or build complex graphs to communicate program semantics to hardware~\cite{zhang_PDGPrefetcherDynamic_2024}. Many rely on some sort of profiling that is then passed down to the hardware to help it issue prefetches~\cite{zhang_RPG2RobustProfileGuided_2024, wang_GuidedRegionPrefetching_2003}, but as mentioned previously, representative profiling is extremely difficult to accomplish in practice. A technique such as Compiler-Directed Content-Aware Prefetching~(\mbox{CDCAP}, described in~\cite{al-sukhni_CompilerdirectedContentawarePrefetching_2003}) is likely most practical, as only minimal yet readily-available static information (LDS node layout) is passed down to hardware.
 
\section{Memory System Trends}
\label{sec:mem_system_trends}

Recent developments in memory technology have seen R\&D budgets utilized more and more towards improving bandwidth, rather than latency, in the hopes of mitigating the off-chip memory bandwidth wall~\cite{rogers_ScalingBandwidthWall_2009}. Examples include  signaling improvements such as three-level (-1, 0, +1) pulse-amplitude modulation~(PAM-3) as used in GDDR7 to transfer 1.5 bits of data per cycle~\cite{yang_131354GbPin_2024}, double data rate interfaces that push data on both edges of a clock~\cite{kalter_50ns16MbDRAM_1990}, and novel quad data rate systems that use two double-pumped clocks at a 90 degree phase offset to transfer \textbf{four} signals per cycle~\cite{jedec_GraphicsDoubleData_2023}. To increase bandwidth, some DRAM chips even utilize additional dimensions of signaling---HBM3 memory, for example, uses 3D-stacked memory dies connected with a silicon interposer to increase link density~\cite{park_192Gb12High896GB_2022, lee_13448GB16High_2024}. This increase in bandwidth reduces the cost of prefetching, as the likelihood of a prefetch causing congestion decreases, and with memory stall times ever-increasing, more aggressive data prefetchers may be the only solution for avoiding memory stalls.

\chapter{Effectively Prefetching Data Structures Linked with Pointers}
\label{chap:design}

We first begin by identifying the following requirements for any effective prefetcher of linked data structures:
\begin{enumerate}
    \item Get \textbf{ahead} of the application, ensuring prefetches are timely (i.e., issued early enough, but not too early~\cite{liu_SemanticsAwareTimelyPrefetching_2010}) so the data is on-chip (ideally in the \mbox{L1-D\$}) when the application needs it. This is especially vital as memory stalls increase.
    \item \textbf{Stay} ahead as execution progresses, continuing to issue prefetches to travel down the LDS by utilizing child pointers found in memory responses.
    \item Be \textbf{accurate} with prefetch requests to avoid cache pollution.
\end{enumerate}

\textsc{Linkey} uses three hardware structures to accomplish these goals. To \textit{get ahead} of the application and ensure prefetches are timely, we add an \textbf{Address Table~(AT)} and \textbf{Child Association Table~(CAT)} that allows us to store the locations and associations between the most important nodes of the data structure. When accesses to roots are detected, these tables allow us to fetch the majority of likely subsequent accesses in parallel. Then, to \textit{stay ahead} of the application, we add a \textbf{Backup Fetch Queue~(BFQ)} to continue fetching after exhausting the data in the tables. We populate this queue using pointers found in memory responses and draw from it when additional memory bandwidth is available for prefetching, e.g., when addresses miss in the tables. 

To ensure prefetches are accurate, \textsc{Linkey} uses metadata provided by the software to only issue prefetches for useful linking pointers. This allows the user/compiler to control which linking pointers should be prefetched for a given traversal. While \textsc{Linkey} cannot predict the next node to be accessed, it can accurately predict the \textit{set} of nodes that will follow the current access, hence why it issues many prefetches at once. The other nodes in the prefetched set are also not useless---there is a (often strong) chance they will be needed soon. Our accuracy and hit count results in \autoref{chap:expirements} reinforce this hypothesis.

A high-level overview of \textsc{Linkey}'s operation follows, along with further details on the organization and maintenance of the AT, CAT, and BFQ\@.

\section{Principle of Operation}
\label{sec:operational_principle}

We chose a hybrid prefetcher design for \textsc{Linkey}. Purely software prefetching approaches are intractable, as low-level information available to the runtime (e.g., pointer values) cannot be statically determined by the compiler. While hardware-only approaches are possible, as evidenced by Intel's CDP~\cite{cooksey_StatelessContentdirectedData_2002}, the lack of high-level information leads to excessive prefetches and cache pollution. Thus, we choose a hybrid approach to utilize the strengths of both the software and the hardware. Specifically, \textsc{Linkey} obtains node layout information from the user/compiler and melds it with pointer values available at runtime to effectively prefetch linked data structures.

A few basic assumptions are also made. We hypothesize that the majority of applications will follow these assumptions, and thus use them as optimization opportunities:
\begin{enumerate}[beginpenalty=10000]
    \item Linked data structures have high reference locality, i.e., child pointers of nodes do not change often.
    \item The majority of traversals of a linked data structure start from a small set of nodes (the ``roots'').  This implies a skewed distribution of node ``temperatures,'' with the roots and nodes near them very hot, while the rest are cold.
    \item A traversal starts when a root is accessed following an access to a different node.
    \item\label{ite:assumptions:keyo} Traversals tend to access non-pointer fields of a node in the same order.
\end{enumerate}

With these assumptions in mind, the \textsc{Linkey} prefetcher is provided with a few small pieces of information: the size of a node in the LDS~(\textit{NodeSize}), the offsets  of linking pointers to other nodes from the start of a node~(\textit{ChildOs}), and the addresses of the ``root'' nodes (stored in the AT and identified with table pointers in a \textit{Roots} register). This information can be provided in many ways; possible implementations include custom instructions, MMIOs, I/O ports, and customized loads/stores. However, the implementation should ideally be serializing with respect to memory, so the prefetcher configuration finishes before memory requests are issued. The serialization can be implemented on the software side with a fence, and will not impact performance as the configuration must only be done once, before any hot loops.

Upon receiving a memory request from the core, \textsc{Linkey} searches the AT for LDS nodes the request address may correspond to. If the address hits, \textsc{Linkey} uses the AT and CAT to issue fetches for the children of the hit node. A check is then performed to see if a new traversal has started; this can also be explicitly indicated with a marker, but doing so is strictly optional. To take advantage of the increased bandwidth available in modern memory systems, as described in \autoref{sec:mem_system_trends}, we allow the prefetcher to output multiple requests per invocation. If the AT/CAT search does not allow the prefetcher to fill its request buffer, further addresses are drawn from the BFQ\@.

When responses arrive from the memory subsystem, the prefetcher is notified and allowed to read the data. If the cache block returned from memory corresponds to any of the addresses in the AT (determined using \textit{NodeSize}), the values of all child pointers in the block (found using \textit{ChildOs}) are extracted and inserted into the AT, then associated with the AT hit using the CAT\@. By using application-provided metadata, we ensure that only pointers known to be useful are added to the tables. Finally, we use a pseudo-LRU eviction system to allocate entries in both tables.

Memory responses are also used to build the BFQ, similar to Intel's CDP system~\cite{cooksey_StatelessContentdirectedData_2002}. When \textsc{Linkey} issues a prefetch, it adds a small amount of metadata to locate the node within the fetched cache block. When the response arrives, the \textsc{Linkey} prefetcher uses this metadata to push all child pointers found in the memory response into the BFQ, excluding those already present in the AT. Then, by pulling from the BFQ when memory bandwidth is available, the prefetcher ``stays ahead'' of the application once it exhausts the entries stored in the AT and CAT\@.

\section{Table-Based Fetching}
\label{sec:table_based_fetching}

\begin{figure}[htbp]
    \centering
    \includegraphics[width=\linewidth,height=0.85\textheight,keepaspectratio]{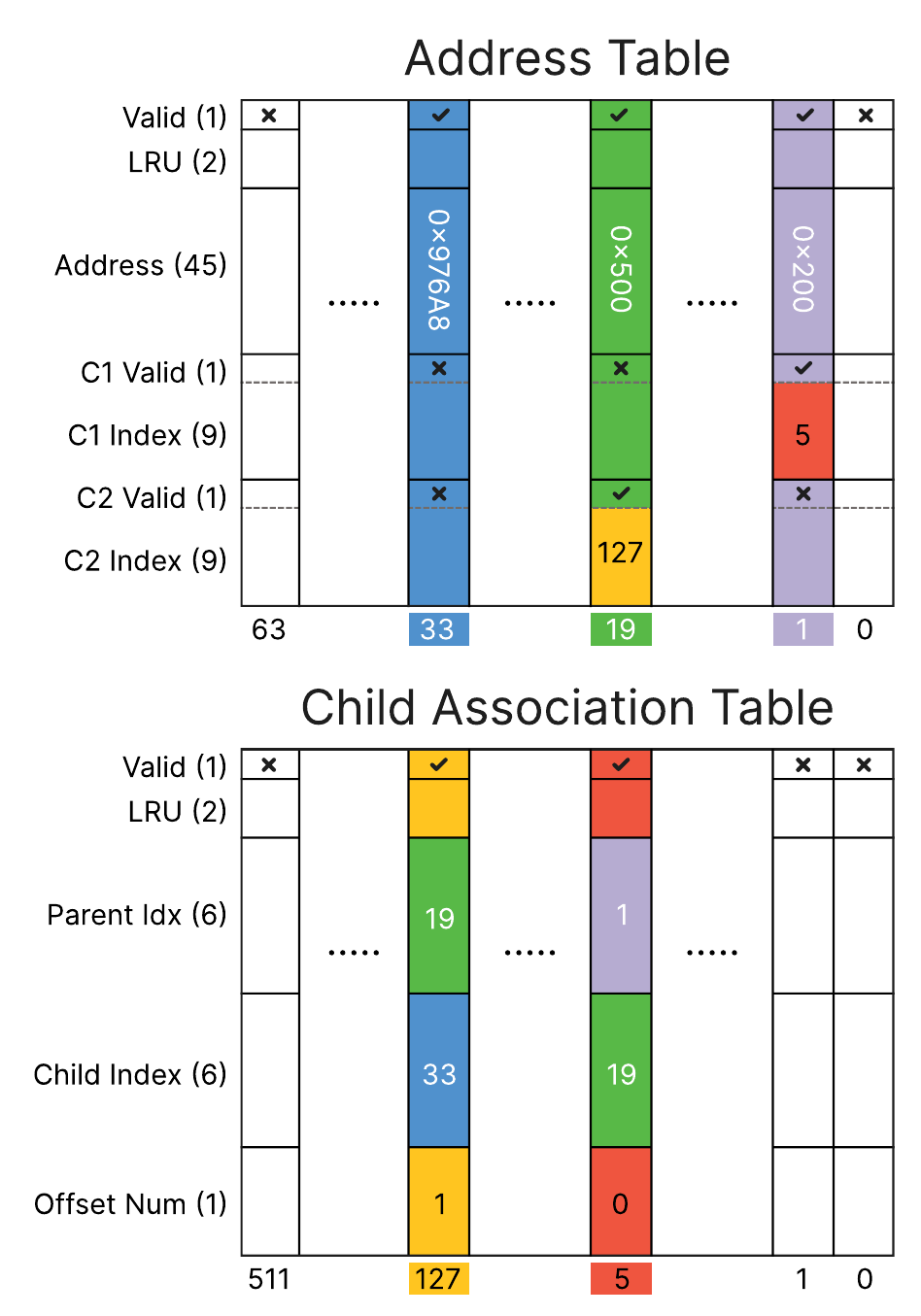}
    \caption[Populated AT and CAT example]{Populated example of the Address and Child Association Tables for a system with 48-bit virtual addresses. The AT holds 64 entries and supports up to two CAT pointers. The CAT holds 512 elements. Note that entries are displayed vertically. The numbers on the bottom of the table represent entry indexes. Colors are used to highlight specific entries.}
    \label{fig:tables}
\end{figure}

We use the AT to keep track of known nodes of the LDS\@. Each AT entry stores a virtual address corresponding to the start of an LDS node, in addition to a valid bit, two LRU bits (described in \autoref{sec:table_evictions}), and the indexes of any child entries in the CAT (each index with its own valid bit).  The CAT associates entries of the AT in a parent/child relationship, with each CAT entry containing a parent and child index, in addition to a valid bit, two LRU bits, and a number specifying which child pointer the entry represents. As nodes of an LDS contain pointers, the nodes will be aligned to the size of a pointer (8 bytes on modern systems), thus, we can elide the bottom 3 bits of the address field in the AT. The AT and CAT organization is further detailed in \autoref{tab:at_entry} and \autoref{tab:cat_entry}, and a populated example of the tables is shown in \autoref{fig:tables}.

\begin{table}[htb]
    \centering
    \begin{tabular}{|c|c|c||c||c||c|}
        \hline
         Valid (1) & LRU (2) & Address & Child 0 Idx/Valid & ... & Child $N$ Idx/Valid \\\hline
    \end{tabular}
    \caption[Address Table entry]{Address Table entry. Bitwidths are shown in parentheses. Address bitwidth depends on machine configuration. Child index bitwidth depends on the number of CAT entries (i.e., $\log_2(|\textit{CAT}\,|)$). $N$ is equal to $|\textit{ChildOs}\,| - 1$.}
    \label{tab:at_entry}
\end{table}

\begin{table}[htb]
    \centering
    \begin{tabular}{|c|c|c|c|c|}
        \hline
         Valid (1) & LRU (2) & Parent Idx & Child Idx & Offset Idx \\\hline
    \end{tabular}
    \caption[Child Association Table entry]{Child Association Table entry. Bitwidths are shown in parentheses. Parent/child index bitwidth depends on the number of AT entries (i.e., $\log_2(|\textit{AT}\,|)$). Offset index bitwidth depends on the number of child pointer offsets (i.e., $\log_2(|\textit{ChildOs}\,|)$.}
    \label{tab:cat_entry}
\end{table}

\subsection{Table Search}
\label{sec:table_search}

When the \textsc{Linkey} prefetcher receives an address from the core, two searches happen in parallel: a root check to identify new traversals, and a full table scan to detect continuing traversals. For the first check, the roots (of which there are a small number) are searched using a base and bound check: $\textit{RootAddr}_i \le \textit{Addr} < \textit{RootAddr}_i + \textit{NodeSize}$. If this check succeeds, we keep the index of the  hit and compute the offset of the request address from the start of the root that hit (i.e., $\textit{Addr} - \textit{RootAddr}_i$). We assume this offset (termed \textit{KeyO}) is the offset of the key used for the traversal (e.g., in a binary tree lookup; see the \ref{ite:assumptions:keyo}th assumption above) and update it when a new traversal begins. If the root check fails, \textsc{Linkey} uses the results of the second check, which is a content-addressable memory~(CAM) lookup along the address field of the table, searching for $\textit{Addr} - \textit{KeyO}$. The use of \textit{KeyO} allows us to avoid the base-and-bound check when searching non-root entries without losing performance. A pseudocode description of this process can be found in \autoref{alg:at_table_search}.

\begin{algorithm}[htbp]
    \caption{AT Table Search}
    \label{alg:at_table_search}

    \begin{algorithmic}[1]
        \LComment{\textnormal{KeyO} is the key offset from the start of the node.}
        \Procedure{SearchRoots}{\textit{Addr}, \textit{NodeSize}, \textit{KeyO}}
            \For {$\textit{Root} \in \textsc{ValidRoots}(AT)$}
                \State $\textit{RA} \gets \textit{Root}.\textit{Address}$
                \If {$\textit{RA} \le \textit{Addr} < \textit{RA} + \textit{NodeSize}$}
                    \If {a new traversal is starting} \Comment{See \autoref{sec:operational_principle} for definition}
                        \State Update \textit{KeyO} to $\textit{Addr} - \textit{RA}$
                    \EndIf
                    \State \Return \textsc{Index}$(\textit{Root}\,)$
                \EndIf
            \EndFor
            \State \Return $\emptyset$
        \EndProcedure
        \\
        \Procedure{CAMTable}{\textit{Addr}, \textit{KeyO}}
            \For {$\textit{Entry} \in \textsc{ValidNonRoots}(\textit{AT}\,)$}
                \If {$\textit{Entry}.\textit{Address} = (\textit{Addr} - \textit{KeyO})$}
                    \State \Return \textsc{Index}$(\textit{Entry})$
                \EndIf
            \EndFor
            \State \Return $\emptyset$
        \EndProcedure
        \\
        \Procedure{SearchAT}{\textit{Addr}, \textit{NodeSize}, \textit{KeyO}} \Comment{Entrypoint}
            \State $I \gets \textsc{SearchRoots}(\textit{Addr}, \textit{NodeSize}, \textit{KeyO})$
            \If {$I = \emptyset$}
                \State $I \gets \textsc{CAMTable}(\textit{Addr}, \textit{KeyO}\,)$
            \EndIf
            \State \Return $I$
        \EndProcedure
    \end{algorithmic}
\end{algorithm}

\subsection{Table Building}
\label{sec:table_building}

Table building occurs upon receiving a memory response for any load, in parallel with the insertion into the cache, or after a store completes. This process runs independently of the table search, and thus can be completely asynchronous. First, a base and bound check is applied to the entire AT, to see which objects might lie in the relevant cache block. For every AT hit, the provided child pointer offsets are used to check if a child pointer may lie in the cache block. If \textsc{Linkey} finds child pointers with non-\texttt{NULL} values within the cache block, it adds them to the AT (if not already present) and allocates new CAT entries to associate the parent AT entry with the child AT entries. If a CAT pointer was already present in the parent AT entry, we first invalidate its corresponding CAT entry. Finally, \textsc{Linkey} stores pointers to the newly created CAT entries in the parent AT entry. This process is detailed in \autoref{alg:table_building}.

\begin{algorithm}[htbp]
    \caption{AT/CAT Table Building}
    \label{alg:table_building}

    \begin{algorithmic}[1]
        \Procedure{SearchBaseBound}{\textit{BlockAddr}, \textit{NodeSize}}
            \State $R \gets \{\}$

            \For {$\textit{Entry} \in \textit{AT}$}
                \State $\textit{SA} \gets \textsc{CacheLine}(\textit{Entry}.\textit{Address})$
                \State $\textit{EA} \gets \textsc{CacheLine}(\textit{Entry}.\textit{Address} + \textit{NodeSize})$
                
                \If{$\textit{SA} \le \textit{BlockAddr} \le \textit{EA}$}
                    \State Add $\textsc{Index}(\textit{Entry})$ to $R$
                \EndIf
            \EndFor

            \State \Return $R$
        \EndProcedure
        \\
        \Procedure{BuildTableForEntry}{\textit{Parent}, \textit{BlockAddr}, \textit{ChildOs}} \label{step:table_building:build_table}
            \For {$o \in \textit{ChildOs}$}
                \State $P \gets \textit{Parent}.\textit{Address} + o$
                \If {$\textsc{CacheLine}(P) = \textit{BlockAddr}$}   
                    \State $C \gets *P$ \Comment{Data is in cache}
                    \If {$\textit{Parent}.\textit{Children}[o].\textit{Valid}$}   
                        \LComment{See \autoref{sec:invalidations} for details on invalidations.}
                        \State $\textsc{InvalidateCATEntry}(\textit{Parent}.\textit{Children}[o].\textit{Index})$ 
                    \EndIf
                    \\
                    \LComment{See \autoref{sec:table_evictions} for details on evictions.}
                    \If {$C \ne \texttt{NULL}$ and \textit{AT} and \textit{CAT} have room after evictions}
                        \State $\textit{Child} \gets \textsc{AddOrFindATEntry}(C)$
                        \State $\textit{CATIdx} \gets \textsc{AddCATEntry}(\textit{Parent}, \textit{Child}, o)$
                        \State \texttt{ }
                        \State $\textit{Parent}.\textit{Children}[o].\textit{Valid} \gets$ \textbf{true}
                        \State $\textit{Parent}.\textit{Children}[o].\textit{Index} \gets \textit{CATIdx}$
                    \EndIf
                \EndIf
            \EndFor
        \EndProcedure
        \\
        \Procedure{BuildTable}{\textit{BlockAddr}, \textit{ChildOs}, \textit{NodeSize}} \Comment{Entrypoint}
            \LComment{Called after a store or when a memory response is received.}
            \State $R \gets \textsc{SearchBaseBound}(\textit{BlockAddr}, \textit{NodeSize})$
            \For {$\textit{Entry} \in R$}    
                \State $\textsc{BuildTableForEntry}(\textit{Entry}, \textit{BlockAddr},  \textit{ChildOs})$
            \EndFor
        \EndProcedure
    \end{algorithmic}
\end{algorithm}

\subsubsection{Evictions}
\label{sec:table_evictions}

We use a modified version of pseudo-LRU to allocate entries in the AT and CAT\@. Two bits are added to each entry: the conventional \texttt{UsedLRU} bit and  a \texttt{JustBuilt} bit. When a table search hits in the AT, we set the \texttt{UsedLRU} bit on the AT entry that hit and any CAT entries that have the AT entry as a parent. The bit is unset when all entries have their \texttt{UsedLRU} bit set. On the other hand, we set the \texttt{JustBuilt} bit during table building, right as we add an entry to the table. \textsc{Linkey} unsets all \texttt{JustBuilt} bits when a new traversal begins (as described in \autoref{sec:operational_principle}). Having either LRU bit set prevents an entry from being evicted; we add the \texttt{JustBuilt} bit to ensure entries aren't evicted before they can be used. Furthermore, a root is \textbf{never} a valid eviction candidate, and neither is the \textit{Parent} entry from the build algorithm (\autoref{alg:table_building}, \autoref{step:table_building:build_table}). Note that this may mean sometimes there are no eviction candidates, in which case we skip insertion. Once an eviction candidate is chosen, \textsc{Linkey} invalidates it (as detailed in \autoref{sec:invalidations}) and then overwrites it.

\subsubsection{Invalidations}
\label{sec:invalidations}

Invalidating a CAT entry is very simple: we mark the corresponding CAT pointer as invalid in the parent AT entry (using the parent index and offset number), and then unset the \texttt{Valid} bit of the CAT entry. To invalidate an AT entry, we invalidate all CAT entries that have that AT entry as either a parent or child (using the CAT invalidation process), then set the \texttt{Valid} bit of the AT entry to 0.

\section{Backup Fetch Queue}

The BFQ is also built asynchronously during the table building phase. When \textsc{Linkey} issues a prefetch, it attaches metadata specifying the offset of the start of the object from the beginning of the requested cache block. If this metadata is present when a block returns from memory, \textsc{Linkey} identifies child pointers with non-\texttt{NULL} values in the cache block using the metadata and \textit{ChildOs}. The node specified by the metadata is also looked up in the AT; this search can be piggybacked off of the table building search in \autoref{sec:table_building}. If the node is not present in the AT, or is present and a child found in the cache block isn't present (which can be checked using the child indexes), \textsc{Linkey} adds that child pointer to the BFQ\@. This last check is to prevent the BFQ from filling with addresses that would already be fetched by the AT and CAT; we assume that a node isn't in the AT if the parent node from the memory response isn't, as the check would otherwise be too expensive.

The prefetch metadata could also be used to avoid the base-and-bound search described in \autoref{sec:table_building} in the majority of cases, however, as the table building runs asynchronously the decision to do so is an implementation detail.

\section{Issuing Requests}

When we prefetch a node, we wish to ensure that the most important portions of the node are brought into the cache. This is extremely important as we allow nodes to span multiple cache blocks, and thus we wish to ensure that only the portions of the node needed for the current traversal are prefetched. To accomplish this goal, we issue prefetches for $\textit{Node.Start} + \textit{KeyO}$ and $\textit{Node.Start} + o$, for all $o \in \textit{ChildOs}$; we know these values will are necessary for the traversal, given our assumptions and the metadata provided by the user. We deduplicate prefetches to ensure the same cache block isn't added to the request buffer twice, and we also prevent prefetches to the cache block specified by the core's demand request.

\begin{figure}[htbp]
    \centering
    \includegraphics[width=\linewidth]{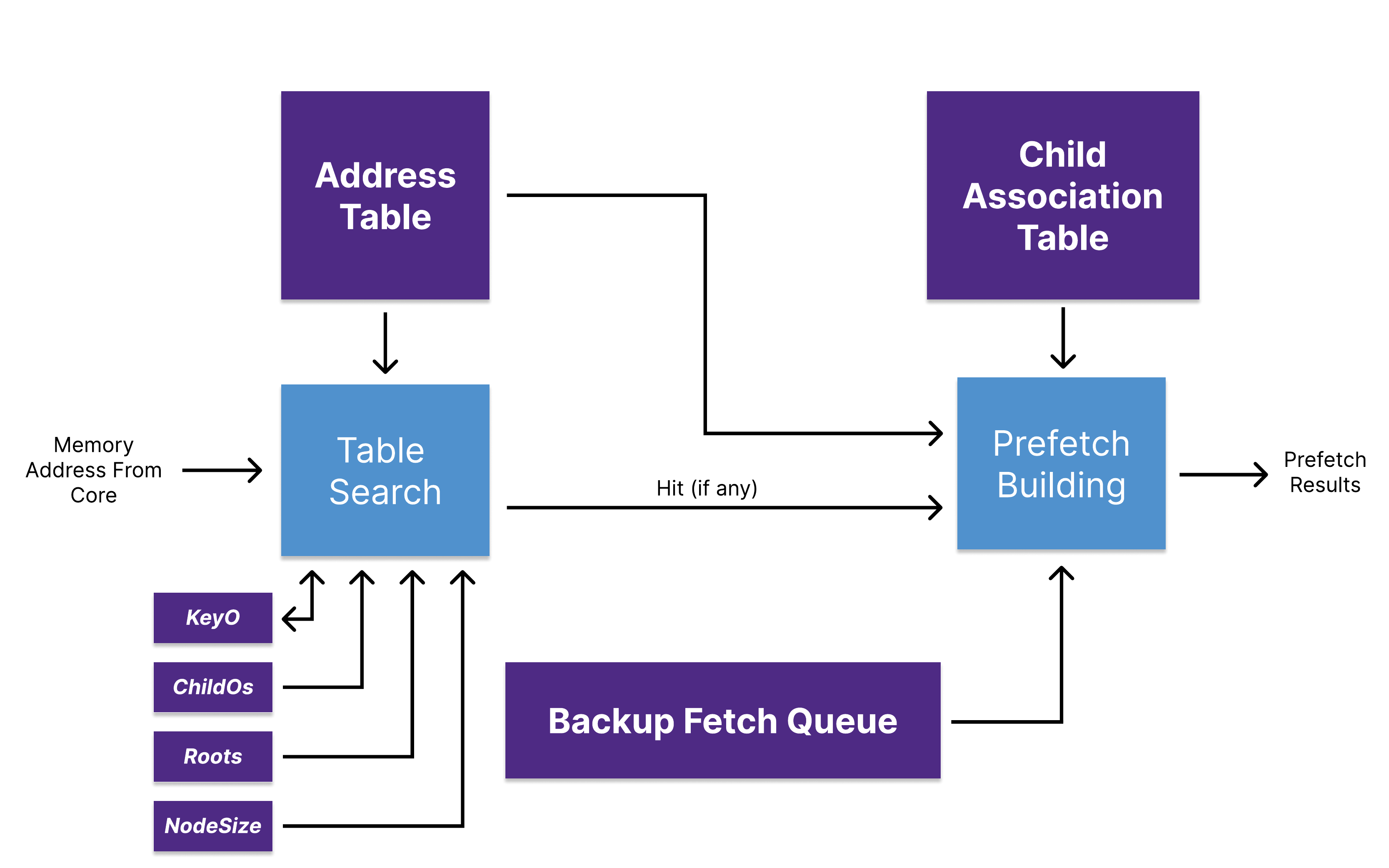}
    \caption[\textsc{Linkey} fetch pipeline]{The \textsc{Linkey} fetch pipeline. This sits on the critical path and runs in parallel to the core's L1-D\$ access.}
    \label{fig:fetch_pipeline}
\end{figure}

The full fetch pipeline is displayed in \autoref{fig:fetch_pipeline}. First, \textsc{Linkey} performs the table search described in \autoref{sec:table_search}. If it finds an AT index matching the address, \textsc{Linkey} adds the index to an internal queue. \textsc{Linkey} then extracts the index from the same queue and issues prefetches for the specified AT entry as described above. The CAT pointers in the AT entry are also used to add the indexes of its child AT entries to the queue. This process continues iteratively, fetching the remainder of the current node and as many child nodes as possible, within the size limit of the output buffer (which we capped at 8 entries). This process allows the prefetcher to ``get ahead'' of the application by fetching, in parallel, the cached shape of the most important nodes of the LDS\@. 

If the output buffer is not full, \textsc{Linkey} draws addresses from the BFQ\@. Prefetches are similarly issued to the $KeyO$ offset and each of the $ChildOs$. As the BFQ is filled using memory responses, employing the BFQ allows the prefetcher to ``stay ahead'' of the application, fetching nodes in the LDS that are not already present in the AT/CAT\@. This entire fetch process is shown in \autoref{alg:req_issuing}.

\begin{algorithm}[htbp]
    \caption{Prefetch Request Issuing}
    \label{alg:req_issuing}

    \begin{algorithmic}[1]
        \LComment{\textit{CoreBlock} is the core memory request's block address}
        \Procedure{IssueRequest}{\textit{Addr}, \textit{ObjectAddr}, \textit{CoreBlock}} 
            \State $\textit{CA} \gets \textsc{CacheLine}(\textit{Addr})$
            \If {$\textit{CA} \ne \textit{CoreBlock}$ and $\textit{CA}$ has not yet been fetched}
                \State $O \gets \textit{ObjectAddr} - \textit{CA}$
                \State Add $(\textit{CA}, O)$ to the request output buffer.
            \EndIf
        \EndProcedure
        \State \texttt{ }
        \Procedure{PrefetchObject}{\textit{BaseAddr}, \textit{ChildOs}, \textit{KeyO}, \textit{CoreBlock}}
            \State $\textsc{IssueRequest}(\textit{BaseAddr} + \textit{KeyO}, \textit{BaseAddr}, \textit{CoreBlock})$
            \For {$o \in \textit{ChildOs}$}
                \If {request buffer is not full}
                    \State $\textsc{IssueRequest}(\textit{BaseAddr} + o, \textit{BaseAddr}, \textit{CoreBlock})$
                \EndIf
            \EndFor
        \EndProcedure
        \State \texttt{ }
        \Procedure{IssueTableFetches}{\textit{EI}, \textit{CoreBlock}, \textit{ChildOs}, \textit{KeyO}}
            \State Initialize an empty queue $Q$
            \State Add \textit{EI} to $Q$
            \State \texttt{ }
            \While {$Q$ is not empty and request buffer is not full}
                \State Pop head of $Q$ into $I$
                \If {$I$ has not yet been seen}
                    \State $\textsc{PrefetchObject}(\textit{AT}\,[I].\textit{Address}, \textit{ChildOs}, \textit{KeyO}, \textit{CoreBlock})$   
                    \State \texttt{ }
                    \For {$c \in \textit{AT}\,[I].\textit{Children}$}
                        \If {$c.\textit{Valid}$}
                            \State $\textit{CI} \gets \textit{CAT}\,[c.\textit{Index}].\textit{ChildIdx}$
                            \State Add \textit{CI} to $Q$
                        \EndIf
                    \EndFor
                \EndIf
            \EndWhile
        \EndProcedure
        \State \texttt{ }
        \Procedure{HandleCoreReq}{\textit{Addr}, \textit{ChildOs}, \textit{KeyO}, \textit{NodeSize}} \Comment{Entrypoint}
            \State $\textit{EI} \gets \textsc{SearchAT}(\textit{Addr}, \textit{NodeSize}, \textit{KeyO})$
            \State $\textit{CoreBlock} \gets \textsc{CacheLine}(\textit{Addr})$
            \If {$\textit{EI} \ne \emptyset$}
                \State $\textsc{IssueTableFetches}(\textit{EI}, \textit{CoreBlock}, \textit{ChildOs}, \textit{KeyO})$
            \EndIf
            \State \texttt{ }
            \While {request buffer is not full}
                \State Pop head of \textit{BFQ} into $A$
                \State $\textsc{PrefetchObject}(A, \textit{ChildOs}, \textit{KeyO}, \textit{CoreBlock})$
            \EndWhile
        \EndProcedure
    \end{algorithmic}
\end{algorithm}

\chapter{Experimental Evaluation}
\label{chap:expirements}

We evaluate \textsc{Linkey} across a wide range of common linked data structures with representative access patterns. We use a modern system configuration as our baseline, and compare performance on key metrics. Finally, we discuss patterns and trends that emerge in the gathered data.

\section{Methodology}

We characterize the performance of \textsc{Linkey} with the Sniper Multicore Simulator~\cite{carlson_SniperExploringLevel_2011, carlson_EvaluationHighLevelMechanistic_2014}, although we only consider single-core performance to simulate the worst-case scenario for LDS applications (i.e., no interleaving of memory requests between threads). We modify the Sniper cache controller to notify the prefetcher when memory responses are returned. Prefetch requests are issued as non-exclusive reads; this is because coherency requests are significantly faster than DRAM as they stay on-chip. Finally, to model systems with higher memory bandwidth, as described in \autoref{sec:mem_system_trends}, we modify the cache controller to issue two prefetch requests at a time and set the prefetch output buffer is set to a size of eight elements. 

We configure the simulator to be representative of a modern high-performance x86-64 system. We adapted simulator configuration parameters from~\cite{cebrian_TemporarilyUnauthorizedStores_2024}; the values are shown in~\autoref{tab:baseline_config}. Virtual and physical address widths were taken from a modern processor (AMD EPYC 7443P). We refer to the baseline \mbox{L1-D\$} striding prefetcher as \texttt{pre\_simple} in our figures.

\begin{table}[h!tbp]
    \centering
    \begin{tabular}{ll}
        \toprule
        ISA & x86-64, 48-bit virtual and physical addresses\\
        Instruction latency (int) & add(1c.), mul(4c.), div(12c.) \\
        \mbox{L1-I\$} & \qty{32}{\kibi\byte}, 8-way, 2-cycle latency \\
        \mbox{L1-D\$} & \qty{48}{\kibi\byte}, 12-way, 5-cycle latency, 64 MSHRs\\
        \mbox{L1-D\$} Prefetcher & Striding, does not cross pages\\
        L2\$ & \qty{1}{\mebi\byte}, 16-way, 16-cycle round trip\\
        L3\$ & \qty{64}{\mebi\byte}, 16-way, 34-cycle round trip\\
        Cache Replacement Policy & LRU for all \\
        DRAM & 160-cycle latency \\
        \bottomrule
    \end{tabular}
    \caption[Baseline system configuration parameters]{Baseline system configuration parameters. We refer to the baseline \mbox{L1-D\$} striding prefetcher as \texttt{pre\_simple} in our figures.}
    \label{tab:baseline_config}
\end{table}

\subsection{Prefetcher Configuration}

To provide basic LDS node layout information to \textsc{Linkey}, as described in \autoref{sec:operational_principle}, we choose to use custom instructions inserted into the program. The added instructions are shown in \autoref{tab:custom_instructions}. To enforce serialization, we insert a \texttt{mfence} into the code after these instructions. However, as the prefetcher configuration only runs once, these instructions are executed outside the hot loop (i.e., the region of interest) of the benchmarks. Additionally, simulator magic instructions are used to view memory response data, as Sniper does not keep track of memory responses; this is purely for simulation within Sniper and would not be needed in a real hardware implementation.

\begin{table}[h!tbp]
    \centering
    \begin{tabular}{lp{11cm}}
        \toprule
        \textbf{Instruction} & \textbf{Description} \\
        \midrule
        \texttt{lds.reset} & Reset the LDS prefetcher and clear all tables. \\
        \texttt{lds.set\_root} & Inform the prefetcher of the address of the $i$th ``root'' node.\\
        \texttt{lds.clear\_roots} & Clear the prefetcher's list of ``root'' nodes. \\
        \texttt{lds.add\_offset} & Inform the prefetcher of a child pointer in the LDS node. \\
        \texttt{lds.set\_size} & Inform the prefetcher of the size of an LDS node (\textit{NodeSize}). \\
        \texttt{lds.new\_traversal} & Inform the prefetcher a new traversal is about to begin, to reset \textit{KeyO} (optional).  \\
        \bottomrule
    \end{tabular}
    \caption[Custom instructions added to interact with \textsc{Linkey}]{Custom instructions added to interact with \textsc{Linkey}.}
    \label{tab:custom_instructions}
\end{table}

We limit the maximum node size (i.e., the distance from the start of the node to the end of the last child pointer) to \qty{4}{\kibi\byte}, giving the \textit{NodeSize}, \textit{KeyO}, and \textit{ChildOs} registers a width of 12 bits. We configure \textsc{Linkey} to support up to eight child pointers, thus \textit{ChildOs} holds eight 12-bit entries. Finally, we allow up to four roots, meaning we must add four registers holding pointers into the AT with corresponding valid bits.

As shown in \autoref{tab:baseline_config}, we assume 48-bit virtual addresses. Thus, we use 45-bit addresses in the AT and BFQ---the lower three bits can be dropped due to structure alignment (see \autoref{sec:table_based_fetching}). The BFQ is limited to eight entries. For the AT and CAT, we test several sizes; the configurations are shown in \autoref{tab:linkey_sizes}.

\begin{table}[h!tbp]
    \centering
    \begin{tabular}{lccc}
        \toprule
        \textbf{Config Name} & \textbf{AT Entries} & \textbf{CAT Entries} & \textbf{HW Size (Bytes)}  \\
        \midrule
        \texttt{pre\_lds\_a64\_c256} & 64 & 256 & \num{1599.5} \\
        \texttt{pre\_lds\_a256\_c1024} & 256 & 1024 & \num{7232.5} \\
        \texttt{pre\_lds\_a1024\_c4096} & 1024 & 4096 & \num{32833.5} \\
        \bottomrule
    \end{tabular}
    \caption[Benchmarked configurations of \textsc{Linkey}]{Benchmarked configurations of \textsc{Linkey}. The size values are totals, including the AT, CAT, BFQ, and all registers (\textit{ChildOs}, \textit{NodeSize}, \textit{KeyO}, and \textit{Roots}).}
    \label{tab:linkey_sizes}
\end{table}

\section{Benchmarks}

We chose our benchmarks to represent a wide range of linked data structures with access patterns representative of a variety of applications. Broadly-speaking, benchmarks can be divided into two categories: \textit{traversal} benchmarks that visit each node in an LDS at least once, and \textit{lookup} benchmarks that search for a certain value in an LDS with the minimum number of node accesses. Additionally, benchmarks can be split into static and dynamic categories, depending on if the LDS is modified during the traversal/search kernel. Descriptions and categorizations of the benchmarks can be found in \autoref{tab:benchmarks}.

\begin{table}[h!tbp]
    \centering
    \begin{tabular}{lp{7.2cm}cc}
        \toprule
        \textbf{Name} & \textbf{Description} & \textbf{Type} & \textbf{Dynamic} \\
        \midrule
        \texttt{ll} & Sum of a linked list. & Traversal & \texttimes\\
        \texttt{ll\_reverse} & Reversal and then sum of a linked list. & Traversal & \checkmark\\
        \texttt{dll} & Sum of a doubly linked list. & Traversal & \texttimes\\
        \texttt{bintree\_dfs} & DFS sum of a binary tree. & Traversal & \texttimes\\
        \texttt{bintree\_bfs} & BFS sum of a binary tree. & Traversal & \texttimes\\
        \texttt{bintree\_probe\_uni} & Probe of a balanced binary search tree with uniformly-distributed random keys. & Lookup & \texttimes\\
        \texttt{bintree\_probe\_zipf} & Probe of a balanced binary search tree with Zipfian-distributed random keys. & Lookup & \texttimes\\
        \texttt{rbtree\_uni} & Probe of a Red-Black Tree with uniformly-distributed random keys. & Lookup & \checkmark\\
        \texttt{rbtree\_zipf} & Probe of a Red-Black Tree with Zipfian-distributed random keys. & Lookup & \checkmark\\
        \texttt{splay\_uni} & Probe of a Splay Tree with uniformly-distributed random keys. & Lookup & \checkmark\\
        \texttt{splay\_zipf} & Probe of a Splay Tree with Zipfian-distributed random keys. & Lookup & \checkmark\\
        \texttt{trie\_uni} & Probe of a trie with uniformly-distributed random keys. & Lookup & \texttimes\\
        \texttt{trie\_zipf} & Probe of a trie with Zipfian-distributed random keys. & Lookup & \texttimes\\
        \texttt{octree} & Traversal of an octree in a W-cycle. & Traversal & \texttimes\\
        \texttt{graph\_bfs} & BFS sum of a graph & Traversal & \texttimes\\
        \bottomrule
    \end{tabular}
    \caption[Benchmarks used to evaluate \textsc{Linkey}]{Benchmarks used to evaluate \textsc{Linkey}.}
    \label{tab:benchmarks}
\end{table}

We run all benchmarks with several input sizes. The \textit{small} size uses $\approx$\num{1000} nodes, the \textit{large} size uses $\approx$\num{10000} nodes, and the \textit{huge} size uses $\approx$\num{100000} nodes. We use a random memory pool to allocate LDS nodes; this is to be representative of real-world applications, as at any point but a cold start, the application's node pool will be arbitrarily ordered. The entire benchmark suite is written in C++ and compiled with GCC~7.3.1 at \texttt{-O3} with inlining disabled.

\subsection{Traversal Benchmarks} 

Approximately half (7 out of 15) of the benchmarks ran are traversal-based benchmarks. These perform common traversals (BFS, DFS, etc.) on linked data structures. As these benchmarks visit each node only one or two times, benefits from \textsc{Linkey} will likely be somewhat limited, as a central assumption of our design is that the distribution of node temperatures is skewed. However, as we are aware of the structure of the LDS, we do still expect to see some improvement.

\subsubsection{Linked-List} 

These benchmarks (\texttt{ll}, \texttt{ll\_reverse}, and \texttt{dll}) all traverse a type of linked list through its entirety, computing a sum along the way. The \texttt{ll\_reverse} benchmark reverses the linked list between summations to evaluate effectiveness on changing data structures. The \texttt{dll} benchmark performs both a forwards and backwards traversal, evaluating effectiveness on data structures with multiple roots and pointer cycles.

\subsubsection{Binary Tree}
\label{sec:bintree_traversal} 

These two benchmarks (\texttt{bintree\_dfs} and \texttt{bintree\_bfs}) perform common traversals of binary tree data structures to compute a sum. The binary tree used consists of $d$ full levels, meaning it is also perfectly balanced. The DFS benchmark is implemented recursively, while the BFS benchmark uses a \texttt{std::queue} from the C++~STL\@.

\subsubsection{Octree} 

The \texttt{octree} benchmark performs a W-cycle traversal on an octree (8-ary tree). This type of traversal is commonly used in Computational Fluid Dynamics algorithms~\cite{khokhlov_FullyThreadedTree_1998, maruszewski_OctreesLatticeBoltzmann_2025} to iteratively step a problem on a coarser/finer mesh, and an octree is specifically used to represent 3-D space. The traversal pattern looks like a recursive ``W'', hence the name. The benchmark kernel is shown in \autoref{lst:octree_w_cycle}.

\begin{lstlisting}[
    language=C,
    float,floatplacement={h!tbp},
    label={lst:octree_w_cycle},
    caption={[Octree ``W-cycle'' traversal kernel]Code to perform the ``W-cycle'' traversal on an octree. Note the double recursion pass.}
]
long sum_tree(node_t* tree) {
    if (tree == NULL)
        return 0;

    // Start with tree value
    long sum = tree->value;

    // Sum children twice (W cycle)
    for (int pass = 0; pass < 2; ++pass) {
        for (int i = 0; i < 8; ++i)
            sum += sum_tree(tree->children[i]);
    }

    return sum;
}
\end{lstlisting}

\subsubsection{Graph} 

The graph benchmark (\texttt{graph\_bfs}) performs a BFS traversal on a connected, undirected graph of max degree five. Nodes have a random number of random children, but must have at least one. All non-\texttt{NULL} children are placed before \texttt{NULL} children. We give \textsc{Linkey} the offsets of the five children of the node, in addition to the full node size and the start node of the traversal. This traversal is implemented using the C++ STL \texttt{queue} and \texttt{unordered\_set}.

\subsection{Lookup Benchmarks}

\begin{figure}
    \centering
    \includegraphics[width=\linewidth]{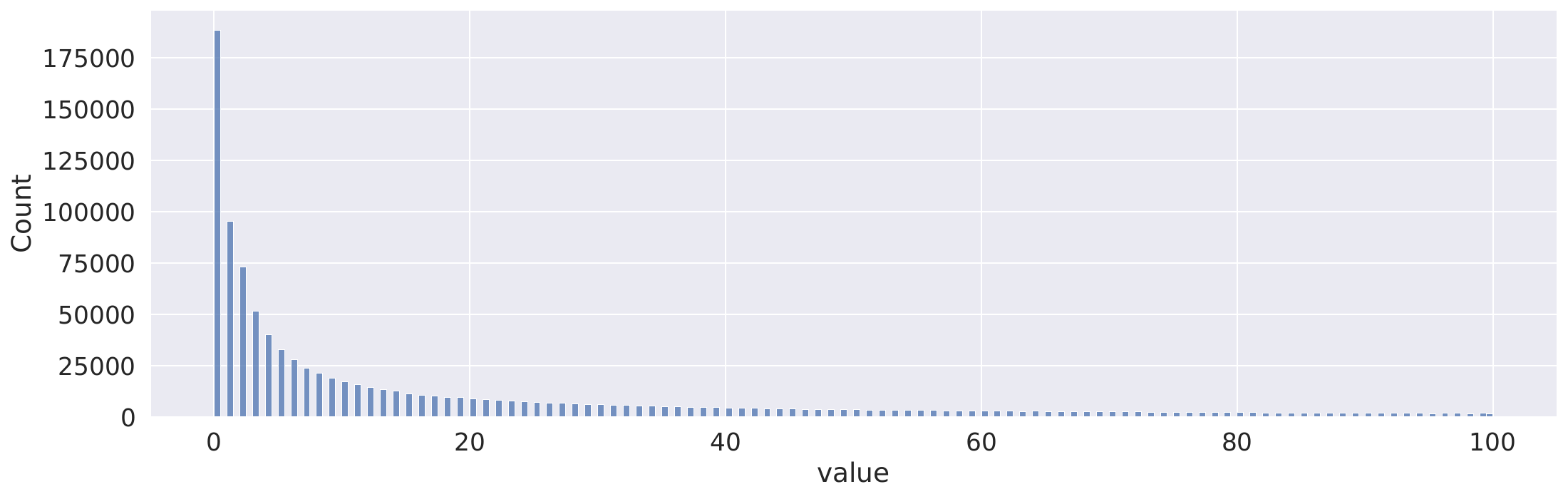}
    \caption[{Zipfian distribution on $[1, 100]$}]{A histogram of a Zipfian distribution with  parameter $\theta=0.99$ on $[1, 100]$. The distribution was sampled \num{1000000} times.}
    \label{fig:zipf_distribution}
\end{figure}

This second category of benchmarks performs many lookups (probes) of a linked data structure. Under the current configuration, we carry out 1000 probes of an LDS of varying size. Keys are randomly generated using either a uniform distribution (\texttt{*\_uni}, implementation from C++ STL) or a Zipfian distribution (\texttt{*\_zipf}, implementation from ~\cite{lersch_LlerschCpp_random_distributions_2024, gray_QuicklyGeneratingBillionrecord_1994}) with a fixed seed. Uniform distributions serve as a worst-case, while Zipfian distributions are more representative of real-world scenarios (a very small number of very hot keys). Our Zipfian distribution used a skew parameter $\theta=0.99$ (taken from YCSB~\cite{cooper_BenchmarkingCloudServing_2010}); a plot of the distribution can be found in~\autoref{fig:zipf_distribution}.

For benchmarks with numerical keys, we generate the keys to range from $1$ to $2^d-1$, where $d$ is the depth of the tree. For the dynamic lookup benchmarks, we set aside 5\% of keys---as in YCSB~\cite{cooper_BenchmarkingCloudServing_2010}---to be randomly inserted during the kernel, rather than during the initial tree build, thus evaluating performance on insertions, changing trees, and missing keys. If the root of the data structure changes (as it might with balanced BSTs), we inform the prefetcher.

We expect to see better performance from these benchmarks compared to the traversal benchmarks, as some nodes will certainly be ``hotter'' than others, especially with Zipfian-distributed keys. However, we do also acknowledge the possibility of a performance hit on dynamic BST benchmarks (i.e., \texttt{rbtree\_*} and \texttt{splay\_*}), as the LDS can change on insertion and even lookup, hurting reference locality.

\subsubsection{Binary Tree} These benchmarks (\texttt{bintree\_probe\_*}) use the same balanced binary search tree as described earlier in \autoref{sec:bintree_traversal}. This tree is static---only lookups are performed.

\subsubsection{Red-Black Tree} The Red-Black Tree~(RBTree)~\cite{guibas_DichromaticFrameworkBalanced_1978} is from of balanced BST\@. This type of tree is widely used in software development---not only is it used throughout Linux~\cite{torvalds_KernelGitTorvalds_2025}, it is the default implementation of \texttt{std::set} and \texttt{std::map} in the C++ STL~\cite{lattner_LlvmLlvmproject_2025, lattner_LLVMCompilationFramework_2004}. This benchmark is dynamic---both insertions and lookups are performed. We adapted the implementation from~\cite{suchy_CARATCaseVirtual_2020}.

\subsubsection{Splay Tree} These benchmarks evaluate performance on Splay Trees~\cite{sleator_SelfadjustingBinarySearch_1985}, another form of balanced BST\@. Splay trees are very interesting because they attempt to optimize for temporal locality by moving recently-used keys (on both insertion \textbf{and} lookup) closer to the root. Overall lookup complexity still remains at $\mathcal{O}(\log n)$. This benchmark is dynamic---both insertions and lookups are performed. We adapted the implementation from~\cite{suchy_CARATCaseVirtual_2020}.

\subsubsection{Trie} A trie~\cite{delabriandais_FileSearchingUsing_1959} is a data structure for efficiently storing strings and searching by prefixes. They have many uses, especially for fast text search~\cite{aho_EfficientStringMatching_1975} and web server routing~\cite{nouvertne_Litestar_2024}. The implementation was adapted from~\cite{vedala_TheAlgorithms_2025}---each node has 26 children (one for each letter of the alphabet), and a boolean marker to signify if the node represents the end of a word. All operations are done in lowercase. A random dictionary is used to populate the tree. As only a maximum of eight offsets can be added to the prefetcher, we chose the eight most common letters in the English language (e, t, a, o, i, n, s, and r). A trie might never change over the lifetime of an application (e.g., in a web server), thus, this benchmark is static---only lookups are performed.

\section{Metrics}

Many statistics are available to judge the efficacy of a hardware optimization. We selected the following four metrics, as they would be most impacted by a prefetcher, to evaluate \textsc{Linkey}:
\begin{description}
    \item[Miss Count] The direct goal of a prefetcher is to reduce the number of cache misses an application experiences. Thus, we look at miss counts to evaluate our performance. Theoretically, reducing the number of cache misses should lead to improved performance.
    \item[IPC] The goal of any hardware optimization is to improve application performance. IPC directly correlates with performance, and we hope to maximize it.
    \item[Prefetch Accuracy] Ideally, prefetchers should only prefetch blocks that will later be used by the application. We would like to minimize the number of unused prefetches, to minimize energy usage and contention on shared resources. However, as \textsc{Linkey} is aggressive to account for high effective memory access times, this statistic will likely suffer.
    \item[Prefetch Hits] Similarly to prefetch accuracy, we would like to view our improvement over baseline on the number of hits to prefetched cache lines. This differs from prefetch accuracy, as it is not weighted by the total number of prefetch requests.
\end{description}

We present miss count and IPC statistics normalized to baseline. We divide the plots into columns by benchmark type (lookup vs.\ traversal). For an aggregate metric, we provide a geomean for both types of benchmark, along with an overall geomean.

\section{Results}

We first begin with an evaluation of different \textsc{Linkey} configurations. Using the results, we choose the configuration that is most reasonable, given performance and hardware resource utilization. We then perform an in-depth evaluation of that configuration, across a wide range of metrics.

\subsection{Configuration Evaluation} 

\begin{figure}
    \centering
    \includegraphics[width=\textwidth,height=0.9\textheight,keepaspectratio]{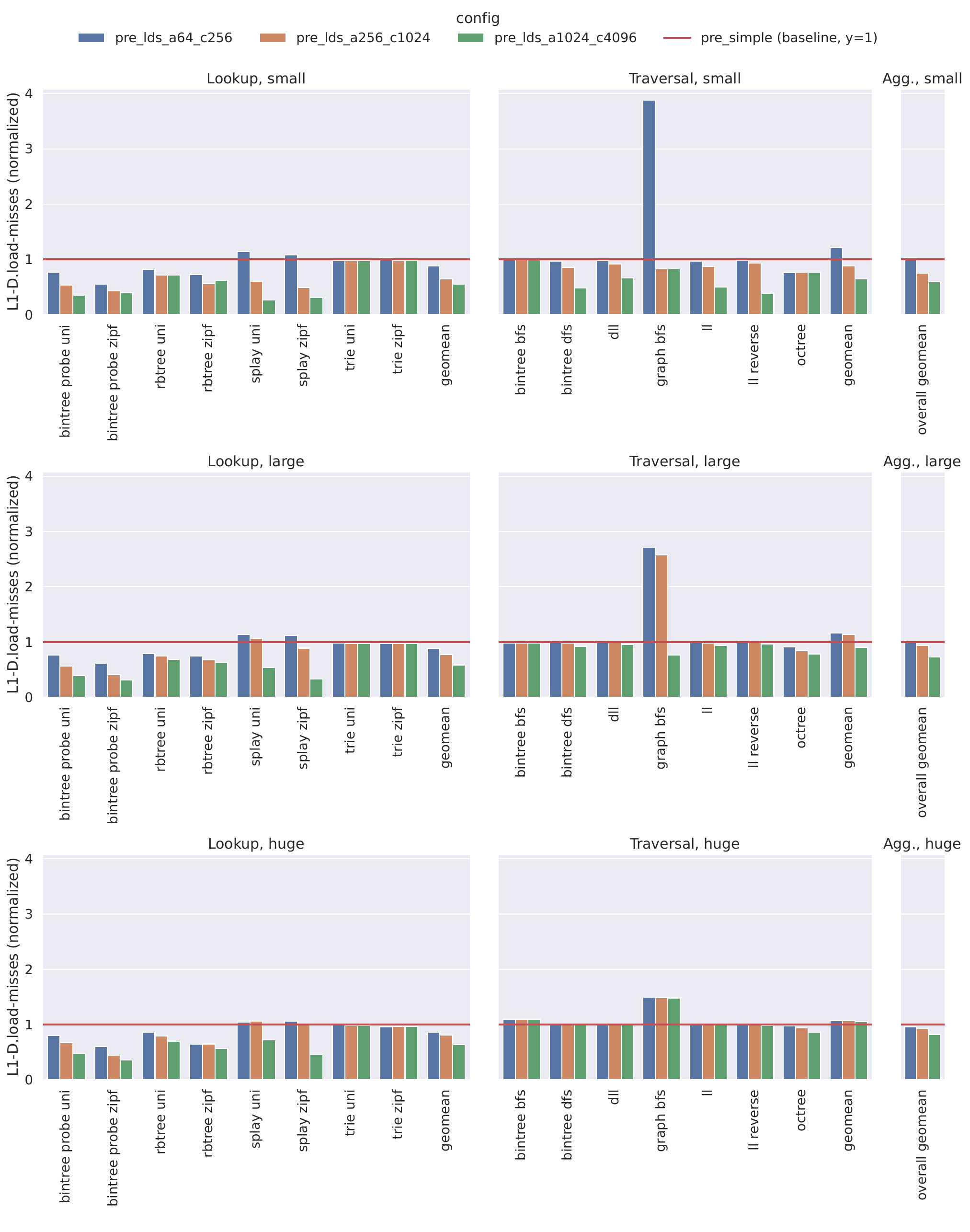}
    \caption[Normalized load misses with different \textsc{Linkey} configurations]{Normalized numbers of load misses with different prefetcher configurations. Rows represent benchmark sizes, colors represent  configurations.} 
    \label{fig:load_misses_size}
\end{figure}

To choose the optimal configuration, we use the two metrics that most directly correlate with performance: miss count and IPC\@. The evaluation of load misses is shown in \autoref{fig:load_misses_size}. The \qty{1.6}{\kibi\byte} version of \textsc{Linkey} is shown in blue, the \qty{7.2}{\kibi\byte} version in orange, the \qty{32.8}{\kibi\byte} version in green, and the striding baseline in red. For readability, we divide the benchmarks by size into rows. Most benchmarks experience a decrease in load misses with the addition of \textsc{Linkey}. At size \textit{huge}, on the bottom,  only \texttt{graph\_bfs} and \texttt{bintree\_bfs} experience increases in the number of load misses. 

\begin{figure}
    \centering
    \includegraphics[width=\textwidth,height=0.925\textheight,keepaspectratio]{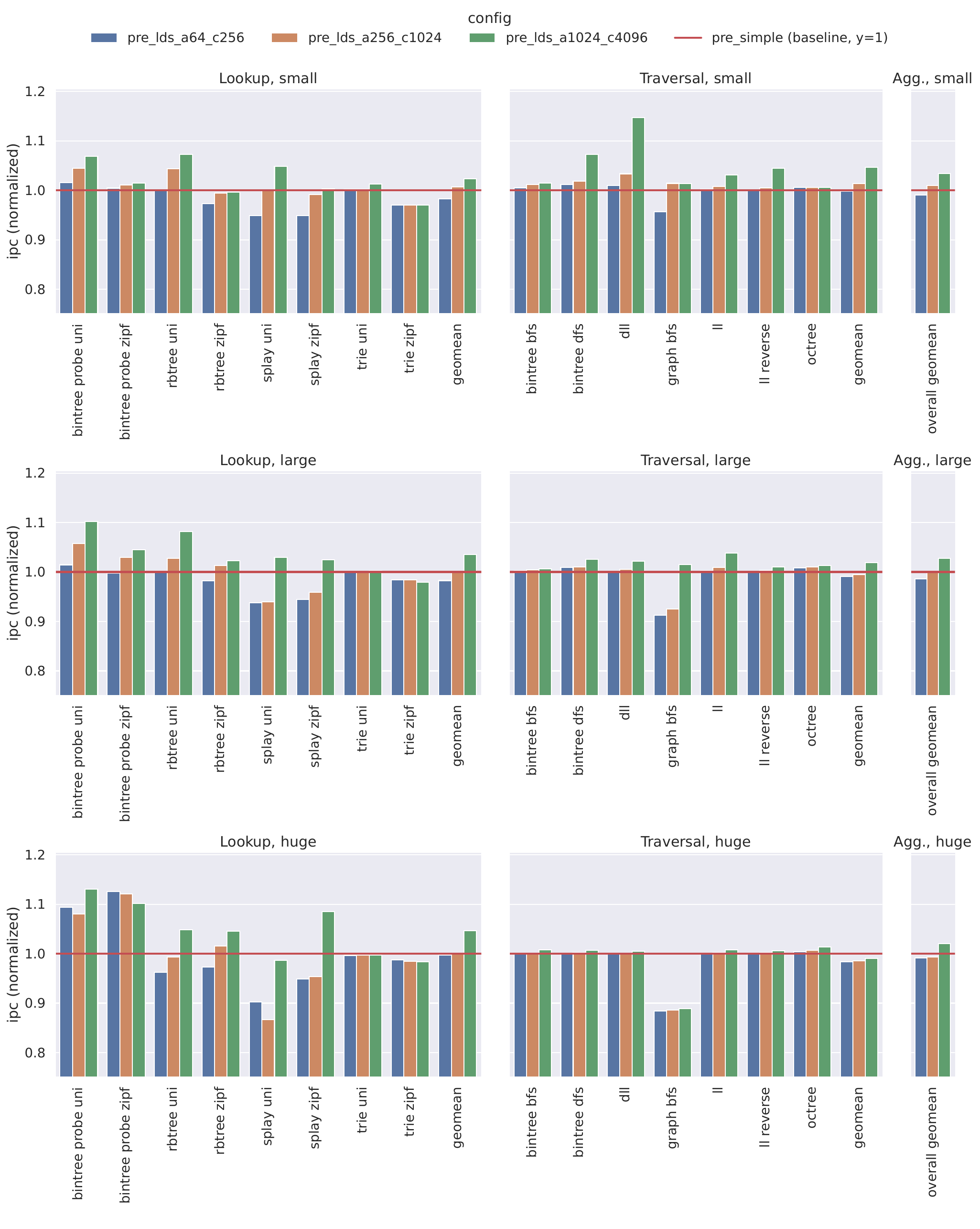}
    \caption[Normalized IPC with different \textsc{Linkey} configurations]{Normalized IPC with different prefetcher configurations. Rows represent  benchmark sizes, colors represent   configurations.}
    \label{fig:ipc_size}
\end{figure}

We show normalized IPC data in \autoref{fig:ipc_size}. All configurations show improvements in IPC on some benchmarks, however, the results are not as consistent as miss count. With the \qty{1.6}{\kibi\byte} configuration, geomean IPC even decreases when the benchmark size is \textit{huge}. However, the \qty{7.2}{\kibi\byte} and \qty{32.8}{\kibi\byte}  configurations do show improvement on most benchmarks at all sizes. 

\subsubsection{Discussion}

The data does show that increased table size leads to improved performance. However, one must consider the scarcity of hardware resources. These graphs show considerable performance improvement with just a 256-entry AT table, and at a size of \qty{7.2}{\kibi\byte} (about 10\% of the L1 caches), the resource utilization is in-line with other modern prefetchers~\cite{peled_NeuralNetworkPrefetcher_2019}. Thus, for the remainder of the evaluation, we will consider only the \qty{7.2}{\kibi\byte} \textsc{Linkey} configuration (\texttt{pre\_lds\_a256\_c1024}).

\subsection{Performance Evaluation}

\begin{figure}
    \centering
    \includegraphics[width=\linewidth]{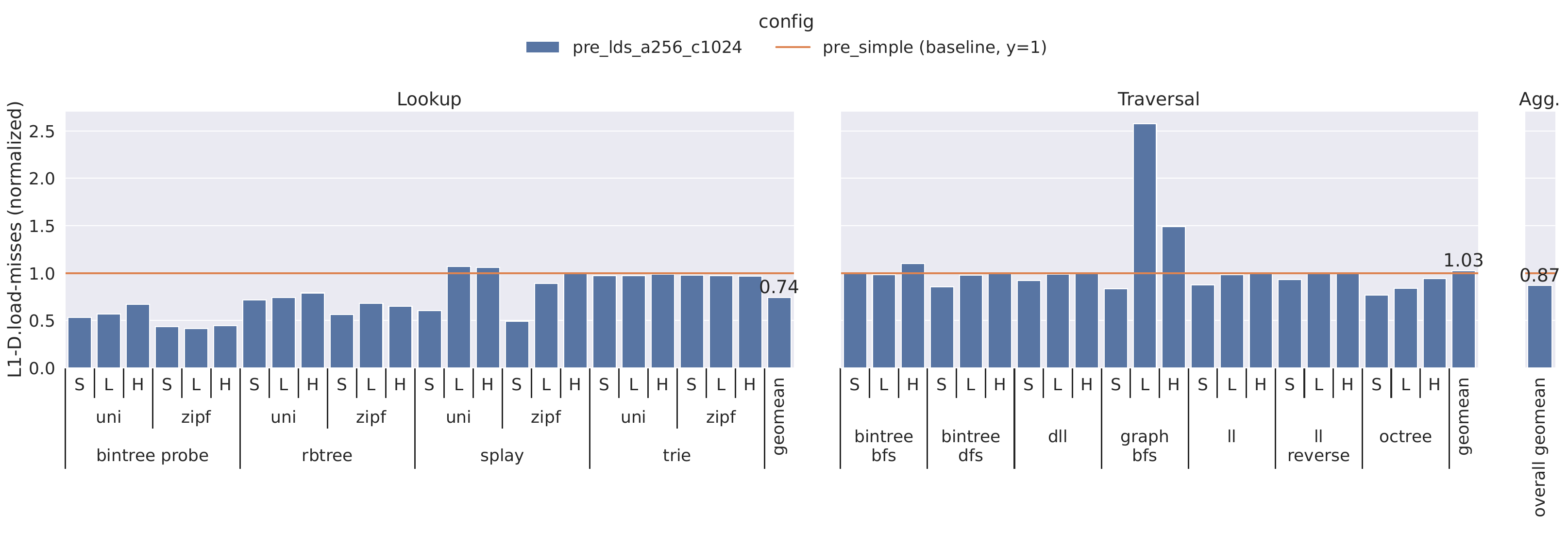}
    \caption[Normalized numbers of load misses]{Normalized numbers of load misses when using the \textsc{Linkey} prefetcher.}
    \label{fig:load_misses}
\end{figure}

In \autoref{fig:load_misses}, we show the normalized load misses between \textsc{Linkey} and the baseline across our benchmark suite. While the Splay Tree and Graph BFS benchmarks see an increase in load misses, we observe a decease in almost all other cases, many times by over $2\times$.  As test sizes increase, load misses tend to increase, although not substantially, demonstrating the scalability of the \textsc{Linkey} approach. Overall, the geomean sees a decrease of 13\%.

\begin{figure}
    \centering
    \includegraphics[width=\linewidth]{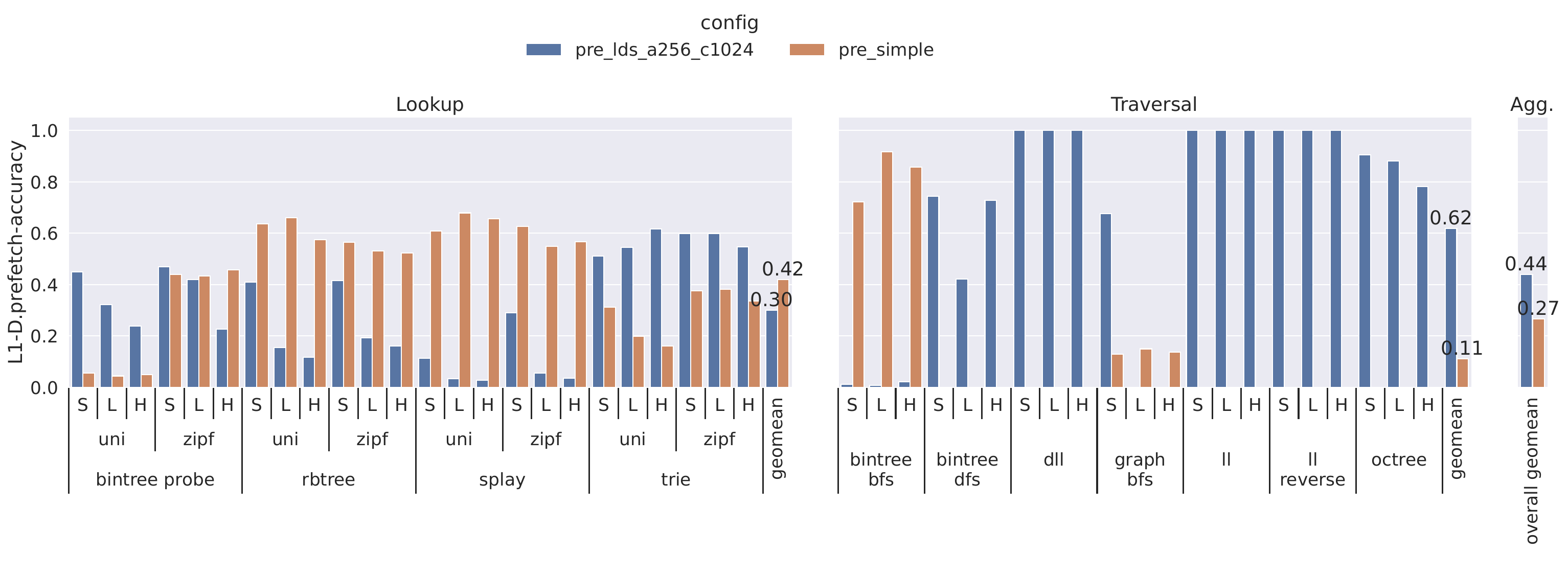}
    \caption[Prefetch accuracy results]{Prefetch accuracy of the \textsc{Linkey} prefetcher and the striding baseline.}
    \label{fig:prefetch_accuracy}
\end{figure}

We show prefetch accuracy results in \autoref{fig:prefetch_accuracy}. Trends in miss rate line up with trends in accuracy, as expected---when accuracy is high, miss rate tends to be lower. Prefetch accuracy also decreases as sizes increase. However, in many cases the baseline had a prefetch accuracy of 0\%---i.e., none of its requests hit at all, while almost all the \textsc{Linkey} prefetches hit in the same benchmarks. While the baseline did do better on some lookup benchmarks, it also issued very few prefetches, meaning the total number of hits was small. Overall, we observe a 65.4\% improvement in mean prefetch accuracy, from 26.6\% to 43.9\%. This shows that \textsc{Linkey} can accurately identify and fetch important child pointers in an LDS\@.

\begin{figure}
    \centering
    \includegraphics[width=\linewidth]{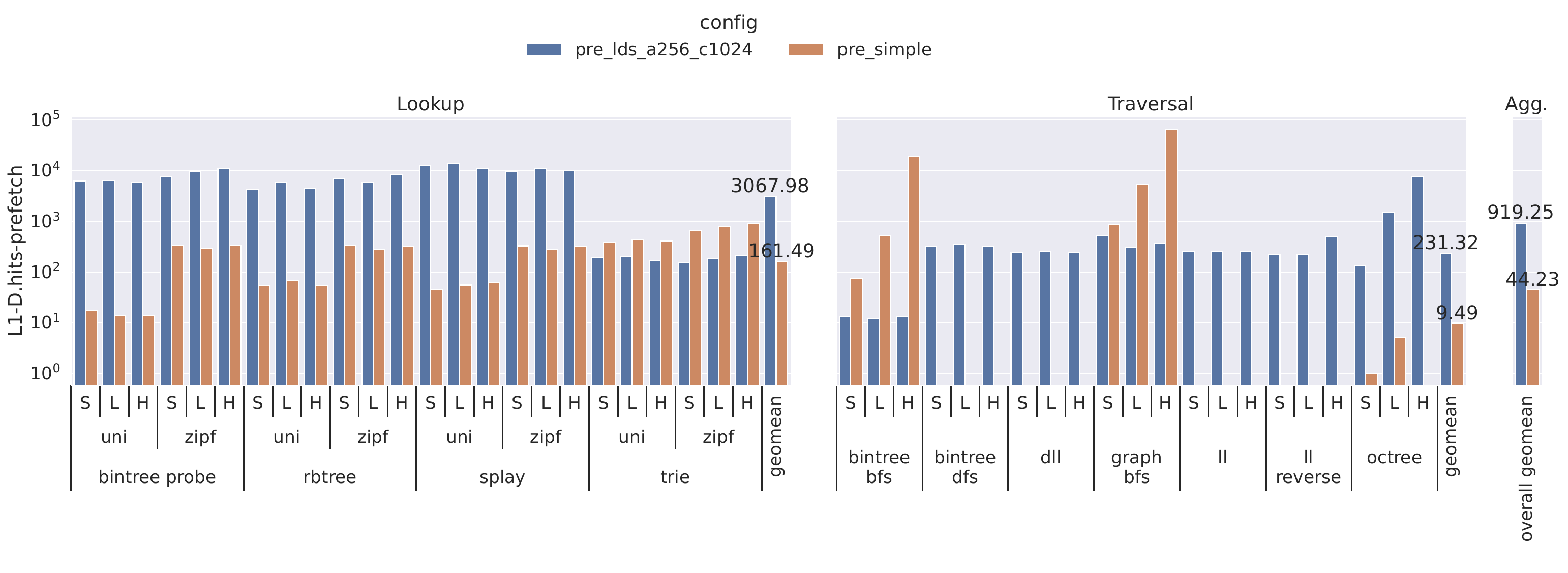}
    \caption[Prefetch hit counts]{Prefetch hit counts of the \textsc{Linkey} prefetcher and striding baseline. Note the use of a log scale on the Y-axis.}
    \label{fig:prefetch_hits}
\end{figure}

We show prefetch hit counts in \autoref{fig:prefetch_hits}. Note the use of a log scale on the Y-axis. In only two benchmarks (\texttt{bintree\_bfs} and \texttt{graph\_bfs}) do we observe significant decreases in the number of prefetch hits---overall numbers increase substantially, sometimes by multiple orders of magnitude, especially on lookup benchmarks. The geomean increase in the number of hits was $19.8\times$; this proves our hypothesis that fetching several ``levels'' of an LDS at a time will improve cache performance. 

\begin{figure}
    \centering
    \includegraphics[width=\linewidth]{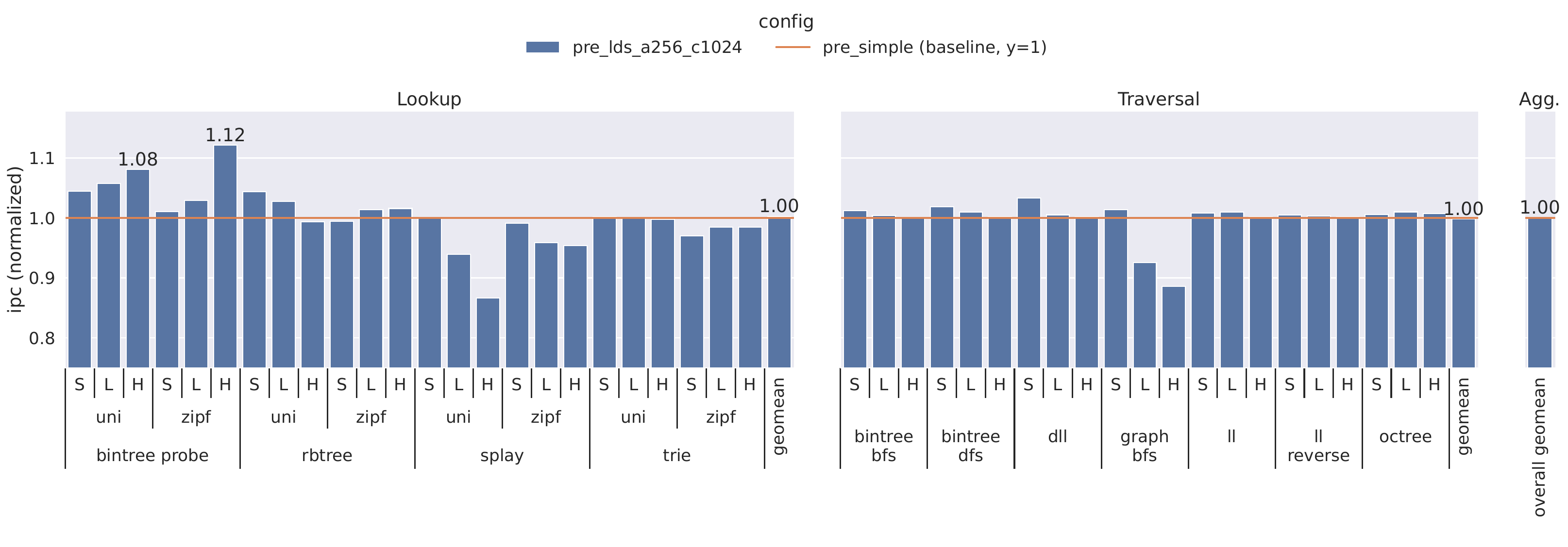}
    \caption[Normalized IPC results]{Normalized IPC when using the \textsc{Linkey} prefetcher.}
    \label{fig:ipc}
\end{figure}

In \autoref{fig:ipc}, we show the IPC results. In cases with low accuracy and high miss counts, we do see some IPC decreases. However, the benchmarks with IPC improvements see substantial improvements---especially \texttt{bintree\_probe\_*} (with the highest improvement of 12.1\%) and \texttt{rbtree\_*}. Additionally, on benchmarks with benefits, IPC gains often improve as test size \textbf{increases}, further demonstrating the scalability of our approach. The overall geomean of normalized IPC increases by 0.05\%.

\subsubsection{Discussion} 

\begin{figure}
    \centering
    \includegraphics[width=\linewidth]{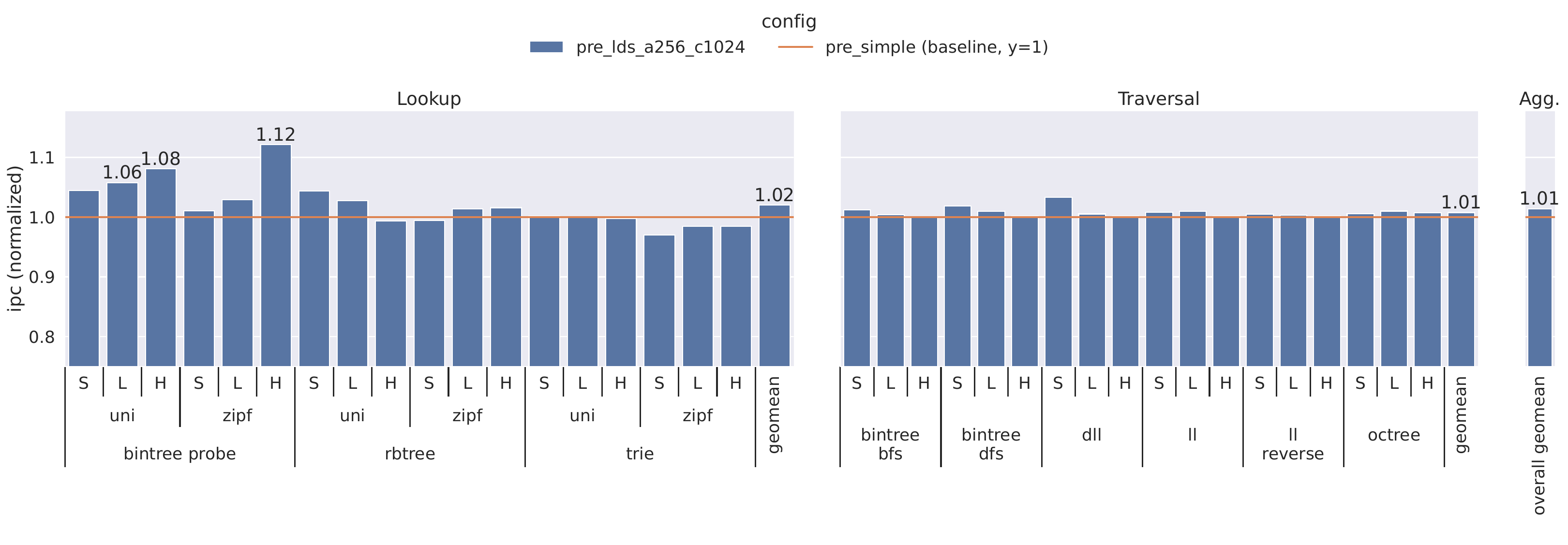}
    \caption[Normalized IPC results, with certain exclusions]{Normalized IPC when using the \textsc{Linkey} prefetcher, with \texttt{graph\_bfs} and Splay Tree benchmarks removed.}
    \label{fig:ipc_better}
\end{figure}

As expected, we see significant performance gains across many benchmarks, especially the lookup benchmarks. While there are IPC decreases on some benchmarks, almost all significant decreases are in the Graph BFS or Splay Tree benchmarks. The performance hit in the graph benchmark is from the nature of BFS---as we visit each node \textit{exactly once}, our assumption that certain nodes are hot breaks down, and the cache ends up polluted with unused prefetches. For splay trees, the constant rearranging leads to \textit{low reference locality}, which predictably  hurts performance due to the need for continuous table invalidations. Our assumption that mutations are significantly rarer than lookups holds with dynamic Red-Black Trees, where we see substantial performance improvements (up to 4.3\% on \texttt{rbtree\_uni-small}). However, as our prefetcher is \textit{configurable}, it can simply be disabled (i.e., just not configured) by the programmer on Splay Trees or graphs traversed in BFS order. IPC with those benchmarks removed is shown in \autoref{fig:ipc_better}, where we observe a geomean performance increase of 1.40\%, and up to 12.1\% on lookup benchmarks.

\chapter{Related Works}
\label{chap:related_works}

With the long history of prefetching technologies, and the importance of linked data structures, there is a significant body of work on algorithms that benefit applications with pointer-chasing access patterns. Here, we focus on the work that best relates to our contributions.

Apple~\cite{vicarte_AuguryUsingData_2022, chen_GoFetchBreakingConstantTime_2024} and Intel~\cite{cooksey_StatelessContentdirectedData_2002} both developed CDPs that identify values in the cache likely to be pointers, with the help of the memory access stream from the core, and issue fetches accordingly. However, this approach has numerous security flaws, as the lack of pointer provenance means data values may be interpreted as pointers and thus fetched, opening a covert channel that attackers can use to read sensitive data through timing attacks. Additionally, these approaches will suffer from degraded performance as effective memory access latencies increase---every cache block must return from memory before it can be scanned for pointers and the next prefetch requests issued. Finally, even if all pointers in a cache block are valid, it is highly unlikely all will be accessed (e.g., cold data pointers), leading to memory system congestion and cache pollution. \textsc{Linkey} avoids these performance issues with its knowledge of LDS node layout.

Efficient Content Directed Prefetching~(ECDP)~\cite{ebrahimi_TechniquesBandwidthefficientPrefetching_2009} builds on Content Directed Pre\-fetching~(CDP) as introduced in~\cite{cooksey_StatelessContentdirectedData_2002} with the goal of eliminating extraneous loads caused by fetching \textit{every} pointer in a cache block. First, the application is profiled, with the goal of finding the offsets of useful loads from the location of the first load in a cache block. When a block is accessed, ECDP searches for values that may be pointers, just like the original CDP, but then utilizes the profiling information to avoid performing certain loads that would not be useful. As a variation of CDP, this technique still falls victim to the issues of sequential loads as detailed above, in addition to the known issues with profiling (e.g., finding fully representative input data). However, it does attempt to solve the extraneous request problem similarly to \textsc{Linkey}---by passing information down from the compiler.

Al-Sukhni et al.~\cite{al-sukhni_CompilerdirectedContentawarePrefetching_2003} introduce Compiler-Directed Content Aware Prefetching~(CDCAP), a technique that also evolves CDP~\cite{cooksey_StatelessContentdirectedData_2002}. Compiler-based profiling is used to statically identify offsets to prefetch, which are then provided to a hardware prefetch engine through special ``Harbinger'' instructions. These instructions are similar to the instructions \textsc{Linkey} uses to convey layout information to the hardware. However, CDCAP only supports a single ``linking'' (i.e., recursive) pointer, meaning performance will suffer on data structures with more than one child. Furthermore, the ``Harbinger'' instructions are inserted into hot loops, requiring compiler support to ``time'' them (using an inherently flawed cost model) and also increasing code size.  Like most approaches, CDCAP must also wait for loads to return from memory before recursing---unlike \textsc{Linkey}, which uses its tables to get ``ahead'' of the application.

In~\cite{wang_GuidedRegionPrefetching_2003}, Wang et al.\ describe a prefetch engine that utilizes hints from a compiler analysis to inform issued requests. Notably, the analysis detects common cases of recursive pointers as used in linked-list traversals, which hardware then exploits to issue prefetches of the LDS up to six levels deep. However, layout information is not provided to the hardware prefetch engine---it instead speculates on values in a cache block being pointers, falling victim to the same cache pollution flaws as other CDPs. Nodes are also assumed to be smaller than two cache blocks, which may not hold across real-world applications. The researchers also state the technique does not perform well on trees. Most importantly, prefetch requests for an LDS are issued sequentially, and thus performance degrades with increased effective memory access times.

Roth et al.\ describe a prefetching technique in~\cite{roth_DependenceBasedPrefetching_1998} that utilizes dependency chains to determine the shape of a linked data structure node in hardware, and issue requests accordingly. While this approach requires no modification of the executable, it cannot learn the full structure of an object without observing accesses to all children. Additionally, the authors limit their prefetcher to only a single node ahead, meaning prefetches are only triggered when the core issues new loads, \textit{and} blocks must be returned from the memory system before new prefetches can be issued---leading to significantly worse performance as effective memory access latencies increase.

Liu et al.~\cite{liu_SemanticsAwareTimelyPrefetching_2010} build on dependence based prefetching~(DBP) as described in~\cite{roth_DependenceBasedPrefetching_1998} to enable \textit{timely} prefetches of children; this is needed as increased memory access times require prefetches to be issued earlier to avoid stalls. Similar to DBP, the instruction and memory access streams are used to build dependency relationships between instructions, however, the authors then classify the loads into direct data accesses, indirect data accesses, and recursive node accesses. This information is then used to issue prefetch loads when memory responses arrive at the core---effectively, this technique is similar to ECDP~\cite{ebrahimi_TechniquesBandwidthefficientPrefetching_2009}, but performs the shape analysis in hardware using methods inspired by DBP\@. As more information is extracted from the instruction stream, more aggressive prefetches can be issued, which helps to improve their ``timeliness.'' However, as the technique attempts to guess if values are pointers,  it is vulnerable to cache pollution like other CDPs---even more so due to the increased aggressiveness of prefetching. Additionally, as the shape analysis is performed in hardware, the full structure of the node can only be learned if accesses to all children are observed. Finally, prefetches are still issued sequentially, meaning with ever-increasing effective memory access times, prefetch timeliness will still suffer. \textsc{Linkey} similarly attempts to improve timeliness, but instead uses its AT/CAT tables to do so.

Jump pointers~\cite{karlsson_PrefetchingTechniqueIrregular_2000, luk_CompilerbasedPrefetchingRecursive_1996, roth_EffectiveJumppointerPrefetching_1999} are a prefetching technique to avoid ``pointer chains'' when prefetching LDSs. Extra pointer(s) are added to each node, referencing the LDS node(s) that will be accessed $N$ timestamps later. Then, the prefetcher can issue requests during a traversal using these jump pointers. This works well if LDS accesses are correlated, however, if something like a tree is being searched based on arbitrary user input, this technique will fail. Additionally, the storage requirements for these pointers present significant overheads, especially on memory-constrained programs.

Correlation prefetchers~\cite{somogyi_SpatiotemporalMemoryStreaming_2009, wenisch_TemporalStreamingShared_2005, joseph_PrefetchingUsingMarkov_1997} check for repeated \textit{streams} of accesses, correlating access patterns with prefetch triggers. They utilize temporal locality to capture sequences of hot access patterns, regardless of complexity. This is very successful on linked data structures with a skewed distribution of node accesses, as the prefetcher can track the hottest streams and issue requests down to the last node. However, the effectiveness of these prefetchers drops as the skewed-ness of an LDS decreases, as only a fixed number of streams can be tracked at once. \textsc{Linkey} takes the opposite approach: rather than issuing very precise fetches to a few nodes, it brings in many nodes that have a high likelihood of being useful in the near future. This allows greater adaptability across access patterns, with the tradeoff that hotter nodes lose a bit of performance so the rest can gain substantially more.

Graph prefetching algorithms have been introduced in~\cite{ainsworth_GraphPrefetchingUsing_2016, kaushik_GretchHardwarePrefetcher_2021, lakshminarayana_SpareRegisterAware_2014, zhang_PDGPrefetcherDynamic_2024} and can significantly improve performance by heavily optimizing for the LDS representations commonly used by graphs. However, these techniques are all so heavily specialized towards a specific application and representation that it is unlikely they will be implemented in a practical system, or have code re-written to fit their programming paradigm.

Many techniques exist that extract a \textit{prefetch kernel}---a small piece of code that mimics the access pattern of the original application---and either assign it to a helper thread running in SMT~\cite{huang_EstimatingEffectivePrefetch_2012, huang_PerformanceAnalysisPrefetching_2009, huang_PerformanceOptimizationThreaded_2012, huidong_PerformanceEvaluationThread_2011}, a small processor in the memory hierarchy~\cite{hughes_PrefetchingLinkedData_2000, yang_ProgrammableMemoryHierarchy_2002, yang_PushVsPull_2000}, or simply embed the kernel in the application code itself~\cite{sankaranarayanan_HelperThreadsCustomized_2020}. While these techniques can produce benefits, the cost of running an entire thread to perform prefetching or redesigning the memory hierarchy to include programmable prefetch engines is too high to be practical. Even just embedding the prefetching kernel in the application code will hurt performance due to code size increases.

Collins et al.~\cite{collins_PointerCacheAssisted_2002} introduce a ``pointer cache'' that breaks serial dependence chains in code such as \texttt{p->next->data}. Effectively, it serves as a value predictor for the \texttt{p->next} load, allowing that load and the load of \texttt{data} to proceed in parallel. The algorithm for determining if a value is a pointer is taken from the CDP paper~\cite{cooksey_StatelessContentdirectedData_2002}. \textsc{Linkey} does not eliminate these dependency chains, but simply attempts to prefetch the pointer data for many nodes in parallel, preventing stalls from occurring.

Finally, while indirect memory prefetchers~\cite{cavus_InformedPrefetchingIndirect_2020, talati_ProdigyImprovingMemory_2021, xue_TycheEfficientGeneral_2024, yu_IMPIndirectMemory_2015} and LDS prefetchers like \textsc{Linkey} have some overlap, the problem spaces are different. Indirect memory prefetchers are typically more focused on array-of-pointers accesses, as would be found in virtual function calls or certain sparse matrix representations. Accesses to these structures do not involve pointer chasing, but instead a (possibly complex) indexing $A[f(B[i])]$. Indirect memory prefetchers attempt to predict the value of $f(B[i])$, a different problem than what \textsc{Linkey} solves.

\chapter{Conclusion and Future work}
\label{chap:conclusion}

Ever-increasing effective memory access times drive the need for novel memory subsystem optimization approaches. Prefetching, while an effective technique for contiguous data structures that follow traditional locality, has struggled when presented with linked data structures. Due to the prevalence and importance of these structures, we must develop new techniques to effectively issue prefetches, or else programs will suffer from ever-increasing memory stalls.

We introduce \textsc{Linkey}, a novel prefetching technique for linked data structures. Using only basic node layout information, \textsc{Linkey} can identify and cache layout information of linked data structures on the core. As memory requests arrive, \textsc{Linkey} both effectively issues prefetches down the LDS and builds its internal representation of the data structure. Unlike other techniques, \textsc{Linkey} does not depend on complex compiler analyses, profiling, or significant user instrumentation to accurately predict future accesses. By allowing the software to specify the link pointers to prefetch, \textsc{Linkey} avoids the cache pollution caused by other CDPs. Furthermore, we designed \textsc{Linkey} with modern hardware in mind, and used increased memory bandwidth available in modern systems to prefetch aggressively in the hopes of avoiding future miss penalties.

With these techniques, \textsc{Linkey} manages to issue targeted prefetches to the most vital nodes of a linked data structure. As a result, we measure substantial performance improvements over a striding baseline. \textsc{Linkey} achieves a geomean 13\% reduction in miss rate with a maximum improvement of 58.8\%, and a 65.4\% geomean increase in accuracy, with many benchmarks improving from 0\%. On benchmarks where \textsc{Linkey} is applicable, we observe a geomean IPC improvement of 1.40\%, up to 12.1\%.

\section{Future Work}

The design and implementation of \textsc{Linkey} could be further extended. We could add more intrusive methods of tracking linked data structures to the hardware (such as pointer tagging to mark provenance), increasing accuracy while remaining fully transparent to the user. We could also perform further benchmarking using other real-world applications, such as the Olden benchmark suite~\cite{carlisle_OldenParallelizingPrograms_1996}. Finally, we could add a separate prefetch buffer to reduce cache pollution, or configure prefetch requests to land in the much larger L2\$---hopefully resolving the slowdowns experienced by the Splay Tree and Graph benchmarks.

\begin{singlespace}
\bibliographystyle{acm} %
\bibliography{LDS-Accelerator}
\end{singlespace}

\appendix		%

\chapter{Other Contributions}
\label{chap:other_contrib}

This prefetching work was performed entirely during my final year at Northwestern University. However, prior to beginning this project, I was working on quantum computing research, specifically on compilation for modular quantum systems. While I carried out the bulk of that work over the prior two years, a substantial portion to wrap up the project and prepare for publication was done during my final year, concurrently with this thesis. The quantum work is currently under review at a top systems conference. The abstract to the paper (which can be found at \href{https://arxiv.org/abs/2501.08478}{arXiv:2501.08478}) is quoted below.

\begin{quote}
As quantum computing technology continues to mature, industry is adopting modular quantum architectures to keep quantum scaling on the projected path and meet performance targets. However, the complexity of chiplet-based quantum devices, coupled with their growing size, presents an imminent scalability challenge for quantum compilation.
Contemporary compilation methods are not well-suited to chiplet architectures. In particular, existing qubit allocation methods are often unable to contend with inter-chiplet links, which don't necessary support a universal basis gate set. 
Furthermore, existing  methods of logical-to-physical qubit placement, swap insertion (routing), unitary synthesis, and/or optimization, are typically not designed for qubit links of wildly varying levels of duration or fidelity.
In this work, we propose SEQC, a complete and parallelized compilation pipeline optimized for chiplet-based quantum computers, 
including several novel methods for qubit placement, qubit routing, and circuit optimization. 
SEQC achieves up to a 36\% increase in circuit fidelity, accompanied by execution time improvements of up to $1.92\times$. Additionally, owing to its ability to parallelize compilation, SEQC achieves consistent solve time improvements of $2-4\times$ over a chiplet-aware Qiskit baseline.
\end{quote}

\section{Personal Contributions}

I was the first person working on this quantum project, and the only one for the first year. My work involved formulating the problem as contained optimization; this led to an initial version of the compiler using MIP solvers. While we achieved good results, the MIP solver was impractically slow, and we pivoted to a greedy approach that later became SEQC\@. My work involved developing and testing initial mapping algorithms, including the Simulated Annealing technique that was utilized in the paper. I also contributed to results collection, analysis, and paper writing. These contributions led to my naming as a \textbf{co-first author} on the final paper.

\begin{vita}                   

Nikola (Nino) Maruszewski is an undergraduate student studying Computer Science at Northwestern University. He is also completing a combined MS in Computer Engineering, advised by Prof. Nikos Hardavellas. Nikola first started his research in quantum computing, where he investigated compilation techniques for modular quantum hardware architectures. However, Nikola now works mainly in the computer architecture space, with his current work in improving prefetching technologies for linked data structures. After his graduation from Northwestern in June 2025, Nikola will be attending Georgia Tech in the fall to pursue a PhD under the supervision of Prof. Josiah Hester.

\end{vita}

\end{document}